\documentclass[npg]{copernicus}
\usepackage{amssymb}
\begin{document}
\newcommand\Alfven{Alfv\'en }
\newcommand{\V}[1]{\mathbf{#1}} 
\newcommand{\bhat}{\mbox{$\hat{\mathbf{b}}$}} 
\newcommand{\figref}[1]{Figure~\ref{#1}}
\newcommand{\secref}[1]{\S\ref{#1}}
\newcommand{\eqref}[1]{equation~(\ref{#1})}
\newcommand{\eqsref}[2]{equations~(\ref{#1})~and~(\ref{#2})}
\newcommand{\om}{\mbox{$\tilde{\omega}$}} 

\title{Limitations of Hall MHD as a model for turbulence in weakly collisional 
plasmas}

\author{Gregory G. Howes}
\affil{505 Van Allen Hall, Department of Physics and Astronomy, University of
Iowa, Iowa City, IA 52242, USA.}

\runningtitle{Limitations of Hall MHD}

\runningauthor{G. G. Howes}

\correspondence{Gregory G. Howes\\ (gregory-howes@uiowa.edu)}

\received{}
\pubdiscuss{} 
\revised{}
\accepted{}
\published{}


\firstpage{1}

\maketitle

\begin{abstract}
The limitations of Hall MHD as a model for turbulence in weakly
collisional plasmas are explored using quantitative comparisons to
Vlasov-Maxwell kinetic theory over a wide range of parameter
space. The validity of Hall MHD in the cold ion limit is shown, but
spurious undamped wave modes exist in Hall MHD when the ion
temperature is finite. It is argued that turbulence in the dissipation
range of the solar wind must be one, or a mixture, of three
electromagnetic wave modes: the parallel whistler, oblique whistler,
or kinetic \Alfven waves. These modes are generally well described by
Hall MHD. Determining the applicability of linear kinetic damping
rates in turbulent plasmas requires a suite of fluid and kinetic
nonlinear numerical simulations.  Contrasting fluid and kinetic
simulations will also shed light on whether the presence of spurious
wave modes alters the nonlinear couplings inherent in turbulence and
will illuminate the turbulent dynamics and energy transfer in the
regime of the characteristic ion kinetic scales.
\end{abstract}

\introduction
Understanding the dynamical evolution of the turbulence in the solar
wind and its thermodynamic consequences for the energy balance in the
heliosphere is a major goal of heliospheric physics. As the time
resolution of \emph{in situ} satellite measurements of this turbulence
is increased, ever smaller length scales of the turbulence are probed.
We have now reached a new frontier in the study of turbulence as we
begin to focus on the dynamics of the plasma turbulence at kinetic
scales---\emph{e.g.}, the ion Larmor radius.  Understanding the
turbulence at these small scales is critical because the kinetic
plasma physics at these scales determines the dissipation of the
turbulent fluctuations and the inevitable conversion of the
fluctuation energy into plasma heat
\citep{Schekochihin:2007,Howes:2008c}.

The simple fluid model of MHD is not sufficient to describe this new
regime of \emph{kinetic turbulence} in the weakly collisional solar
wind plasma.  The appealing simplicity of the Hall MHD model---an
extension of MHD that maintains a fluid description while accounting for
the Hall terms in Ohm's Law, leading to dispersive behavior at small
scales---has lead a number of investigators to use it to study
turbulence at these small scale
lengths \citep{Shebalin:1991,Huba:1994,
Ghosh:1996,Sahraoui:2003,Krishan:2004,Hori:2005,Mininni:2005,Dmitruk:2006,
Servidio:2007}, with some recommending Hall MHD as the
preferred model for the study of solar wind turbulence
\citep{Matthaeus:2008a}. The applicability of the Hall MHD model to 
turbulence in a weakly collisional plasma such as the solar wind,
however, has been called into question \citep{Howes:2008d}.  Here we
investigate the limitations of the Hall MHD model for the study of
turbulence in kinetic plasmas.

Hall MHD is a rigorous limit of the kinetic behavior in a weakly
collisional plasma only if all of the following conditions are
satisfied: $T_i \ll T_e$, and $k_\parallel v_{ti} \ll \omega \ll
k_\parallel v_{te}$ \citep{Ito:2004}. Although this \emph{cold ion
limit} is not universally applicable to the turbulent solar wind, we
nevertheless would like to estimate quantitatively the error in using
the fluid description given by Hall MHD rather than a kinetic
description.  Although a turbulent plasma inherently involves
dynamically significant nonlinear wave-wave interactions which drive
the turbulent cascade of energy to small scales, we focus our
attention on the ability of compressible Hall MHD to reproduce the
properties of the linear wave modes of the Vlasov-Maxwell kinetic
theory. If the linear physics described by Hall MHD deviates from the
kinetic description, then it is likely that the nonlinear wave-wave
couplings that comprise the turbulence will be incorrect as well.

A comparison of the (complex) linear eigenfrequencies determined by
compressible Hall MHD and Vlasov-Maxwell kinetic theory provides a
simple quantitative measure of the fidelity of the Hall MHD
description to the kinetic physics in a weakly collisional plasma.
\citet{Krauss-Varban:1994} have already published a very thorough 
comparison of the low-frequency wave mode properties of Hall MHD and
kinetic theory, concluding that ``fluid theory does not correctly
describe the mode structure and mode properties for most plasmas of
interest in space physics.''  Their study, however, evaluated the
properties at a particular wavenumber $k d_i = 0.1$, where the $d_i$
is ion inertial length---see Table~\ref{tab:defs} for all symbol
definitions. Because the turbulent cascade extends over a large range
of scales (more than four orders of magnitude in the solar wind), the
accuracy of Hall MHD in describing the turbulence must be assessed
over a logarithmic scale in wave vector space $(k_\perp,k_\parallel)$.

This paper aims to identify the regimes of parameter space for which
Hall MHD is not an accurate description. The Hall MHD and
Vlasov-Maxwell systems are specified, and their properties briefly
explored, in \secref{sec:hmhd} and \secref{sec:vm}. In
\secref{sec:comp}, some of the subtleties involved in quantitatively comparing 
Hall MHD to kinetic theory are discussed and the measures of comparison
are defined. The results of the comparison are presented in 
\secref{sec:results}, the implications of choosing the Hall MHD model 
to describe kinetic turbulence are discussed in \secref{sec:disc},
and the key findings are summarized in \secref{sec:conc}.

\section{Standard Hall MHD}
\label{sec:hmhd}
Consider a fully ionized, homogeneous plasma of protons and electrons
threaded by a straight, uniform magnetic field $\V{B}_0$.  By
retaining the $\V{J} \times \V{B}$ Hall term in Ohm's law during the
derivation of the MHD equations, the following form for the
compressible Hall MHD equations is obtained:
\begin{equation}
\frac{\partial \rho}{\partial t} +\nabla \cdot (\rho  \V{u}) =0
\end{equation}
\begin{equation}
\frac{\partial \V{u}}{\partial t} + \V{u} \cdot \nabla \V{u}= 
- \frac{1}{\rho} \nabla p + \frac{1}{4 \pi \rho}(\nabla \times \V{B}) \times \V{B}
\label{eq:hmhd_mom}
\end{equation}
\begin{equation}
\frac{\partial \V{B}}{\partial t}= \nabla \times \left[\V{u} \times \V{B} - 
\frac{c}{4 \pi \rho} \frac{m_i}{q_i} (\nabla \times \V{B})  \times \V{B} \right]
\label{eq:induction}
\end{equation}
\begin{equation}
\frac{d}{d t} \left(\frac{p}{\rho^\gamma}\right)=0
\label{eq:eqstate}
\end{equation}
where the substantial time derivative is $d/d t =\partial/\partial t +
\V{u} \cdot \nabla $ and other terms are defined in
Table~\ref{tab:defs}.  Although there exist numerous variations of
Hall MHD---\emph{e.g.}, employing a double adiabatic equation of
state---this is the simplest compressible system incorporating the
Hall effect, and generally appears to be the most widely used in
studies employing Hall MHD
\citep{Lighthill:1960,Kuvshinov:1994,Mininni:2002,Ohsaki:2004,Hirose:2004,Servidio:2007}.
We therefore refer to this particular system as \emph{standard} Hall
MHD.
\begin{table}[t]
\caption{Definitions}
\vskip4mm
\centering
\begin{tabular}{lc}
\tophline
Variable & Symbol \\
\middlehline
Species & $s$ \\
Speed of light & $c$ \\
Mass & $m_s$ \\
Charge & $q_s$ \\
Number Density & $n_s$ \\
Temperature & $T_s$ \\
Species Plasma Beta & $\beta_s = 8 \pi n_s T_s /B^2$ \\
Plasma Frequency & $\omega_{ps}= \sqrt{ 4 \pi n_s q_s^2 /m_s}$\\
Cyclotron Frequency & $\Omega_s = q_s B/(m_s c)$ \\
Thermal Velocity & $v_{{\rm t}s} =  \sqrt{2 T_s/m_s}$ \\
\Alfven Velocity & $v_A= B/\sqrt{4 \pi n_i m_i}$ \\
Ion Larmor Radius & $\rho_i = v_{{\rm t}i}/\Omega_i$ \\
Ion Inertial Length & $d_i= c/\omega_{pi} = \rho_i/\sqrt{\beta_i}$ \\
\bottomhline
\end{tabular}
\label{tab:defs}
\end{table}

The linear dispersion relation for standard Hall MHD may be written in 
dimensionless form as
\begin{eqnarray}
\lefteqn{\left[\om^2- (k_\parallel d_i)^2\right] \left[\om^4 - \om^2 (k d_i)^2 ( 1 + \beta_0)
+  (k d_i)^2 (k_\parallel d_i)^2 \beta_0 \right] } & &  \nonumber \\
& = & \om^2 (k d_i)^2 (k_\parallel d_i)^2 \left[ \om^2- (k d_i)^2 \beta_0\right]
\label{eq:hmhd}
\end{eqnarray}
where the frequency is normalized to the ion cyclotron frequency $\om
=\omega/\Omega_i$, the plasma beta is $\beta_0 = c_s^2/v_A^2$ with the
sound speed defined by $c_s^2 = \gamma p/\rho$.  It is clear from the
form of this dispersion relation that the eigenfrequencies depend 
on only three dimensionless parameters: the normalized components of the
wave vector with respect to the mean magnetic field, $k_\perp d_i$ and
$k_\parallel d_i$, and the plasma beta $\beta_0$. Thus,
$\omega=\omega_h (k_\perp d_i, k_\parallel
d_i,\beta_0)$.  

The six solutions of the standard Hall MHD dispersion
relation~(\ref{eq:hmhd}) correspond to three distinct physical wave
modes, each with two counter propagating solutions. \footnote{Already
factored out from \eqref{eq:hmhd} is a non-propagating, $\omega=0$
entropy mode, corresponding to pressure balance but a non-zero entropy
fluctuation. This mode, examined in \citet{Hameiri:2005}, is not
discussed in this paper.} In the MHD limit $k d_i \ll 1$, the three solutions to
\eqref{eq:hmhd} give the usual fast, \Alfven, and slow wave
eigenmodes. We will refer to the continuation of each these modes at
$k d_i \gtrsim 1$ by their name in the MHD limit: thus, whistler waves
correspond to the fast wave branch, and kinetic \Alfven waves and ion
cyclotron waves are the extension of the \Alfven branch into the
nearly perpendicular and nearly parallel kinetic regimes,
respectively.

\citet{Hirose:2004} and \citet{Ito:2004} found that the Hall MHD 
dispersion relation~(\ref{eq:hmhd}) is a rigorous limit of
Vlasov-Maxwell kinetic theory if
\begin{equation}
\omega  \ll  \Omega_i  
\label{eq:omllomi} 
\end{equation}
and only if
\begin{eqnarray}
T_i &\ll& T_e \label{eq:tll1} \\
\omega & \gg & k_\parallel v_{ti} \label{eq:omggi}    \\
\omega & \ll & k_\parallel v_{te} \label{eq:omlle}.
\end{eqnarray}
Although condition~(\ref{eq:omllomi}) is sufficient to recover the
kinetic results, it is not a necessary condition under all
circumstances. In a study of the fluid closures for collisionless
plasmas, \citet{Chust:2006} suggested two constraints for validity:
(a) appropriate modeling of the pressure term in the momentum
equation~(\ref{eq:hmhd_mom}); and (b) the negligibility of the
pressure gradient term in Ohm's law. They showed a sufficient
condition satisfying the second constraint under general conditions is
$(k \rho_i)^2 \ll \omega/\Omega_i$.
For the isotropic equation of state~(\ref{eq:eqstate}) used in
standard Hall MHD, the scalar treatment of the pressure means that the
pressure gradient term in Ohm's law is always negligible because Ohm's
law enters the induction equation~(\ref{eq:induction}) only through
its curl. 

A complementary calculation shows that gyrokinetics
\citep{Rutherford:1968,Frieman:1982,Howes:2006}, a rigorous anisotropic limit
of kinetic theory for $k_\parallel \ll k_\perp$ and $\omega \ll
\Omega_i $, yields a reduced (anisotropic) version of Hall MHD in the limit 
$T_i \ll T_e$ \citep{Schekochihin:2007}.  The description of a
collisionless plasma by a fluid model, such as Hall MHD, is only valid when wave-particle
interactions and finite-Larmor-radius effects are negligible
\citep{Ballai:2002}; this is ensured by the
conditions~(\ref{eq:omllomi})--(\ref{eq:omlle})

Condition~(\ref{eq:tll1}) is the \emph{cold ion approximation},
meaning that the ion temperature is negligible compared to the
electron temperature.  In this limit, the sound speed $c_s$ in the
standard Hall MHD system is the ion acoustic speed
$c_s^2=T_e/m_i$.  Condition~(\ref{eq:omggi}) means that the ion Landau
resonance is negligible because the ions are too cold, while
condition~(\ref{eq:omlle}) means that the electron Landau resonance is
negligible because the electrons are too hot. Implied by condition~(\ref{eq:tll1}),
but not explicitly stated in \citet{Ito:2004}, is the requirement that
$k_\perp \rho_i \ll 1$ \citep{Ballai:2002}; the ions must be cold
enough such that the ion Larmor radius is small compared to the
perpendicular scales of interest, and therefore
finite-ion-Larmor-radius effects may be  neglected. The ion acoustic
Larmor radius $\rho_s= c_s/\Omega_i$, however, may be comparable to
the perpendicular scale, $k_\perp \rho_s \gtrsim 1$.

Although few weakly collisional plasmas in space or astrophysical
environments satisfy these very restrictive
conditions~(\ref{eq:omllomi})--(\ref{eq:omlle}), one may ask at what
point will non-negligible kinetic effects lead to a poor description
by Hall MHD.  In this spirit, we wish to evaluate in this paper the
practical limitations of the Hall MHD model for turbulence in a weakly
collisional plasma, making quantitative comparisons of standard Hall
MHD with Vlasov-Maxwell kinetic theory in the parameter regime
relevant to turbulent solar wind.
\section{Vlasov-Maxwell Kinetic Theory}
\label{sec:vm}
To test the Hall MHD model, we compare the real frequency from its
dispersion relation to the complex eigenfrequencies from the
Vlasov-Maxwell dispersion relation
\citep{Stix:1992} assuming a homogeneous, fully ionized proton and electron plasma 
with isotropic Maxwellian equilibrium distribution functions and no
drift velocities. This dispersion relation is solved numerically (see
\citet{Quataert:1998} for a description of the code used to solve it), performing 
the Bessel function sums to 100~terms to ensure accurate results at
high $k_\perp d_i$. The linear Vlasov-Maxwell dispersion relation
depends on five parameters: the normalized perpendicular wavenumber
$k_\perp d_i$, the normalized parallel wavenumber $k_\parallel
d_i$, the ion plasma beta $\beta_i$, the ion to electron temperature
ratio $T_i/T_e$, and the ratio of ion thermal velocity to the speed of
light $v_{ti}/c$. The solution may then be expressed as
$\omega=\omega_v (k_\perp d_i, k_\parallel
d_i,\beta_i,T_i/T_e,v_{ti}/c)$.  A realistic ratio of proton to
electron mass, $m_i/m_e=1836$, is used. We note here that, in the
non-relativistic limit $v_{ti} \ll c$ relevant to the turbulent solar
wind, the low-frequency modes are rather insensitive to the specific
value of $v_{ti}/c$, so we choose the typical value $v_{ti}/c=10^{-4}$
for all comparisons in this paper.  Just as in the Hall MHD case, we
choose to label to the low-frequency wave mode solutions (low compared to 
the electron cyclotron frequency  $\Omega_e$) of Vlasov-Maxwell kinetic theory by their names in the
MHD limit---the fast, \Alfven, and slow waves.

\section{Comparing Hall MHD to Kinetic Theory}
\label{sec:comp}
\subsection{Subtle Aspects of the Comparison}
\label{sec:subtle}

Two complications arise when making a detailed comparison of standard
Hall MHD with Vlasov-Maxwell kinetic theory.  The first is the use of
an isotropic equation of state~(\ref{eq:eqstate}) in standard Hall
MHD. Condition~(\ref{eq:omlle}) suggests that the electron response is
so fast that, if the ion temperature is ignorable, the pressure
response may be correctly treated as isothermal and isotropic
\citep{Hirose:2004}. But for finite ion temperatures, the ion response
is anisotropic and appears to be better described with a double
polytropic equation of state,
\[
\frac{d}{dt}\left( \frac{p_\perp}{\rho B^{\gamma_\perp-1}} \right)=0 \quad
\mbox{ and } \quad 
\frac{d}{dt}\left( \frac{p_\parallel B^{\gamma_\parallel-1}}{\rho^{\gamma_\parallel}} \right)=0. 
\]
For example, a choice of $\gamma_{\perp}=2$ and
$\gamma_{\parallel}=3$ corresponds to the double adiabatic, CGL
closure for the ions
\citep{Chew:1956,Abraham-Shrauner:1973,Ballai:2002}.
Even this more complicated closure, however, fails to yield a
dispersion relation that is a rigorous limit of kinetic theory for a
weakly collisional plasma with a finite ion temperature
\citep{Hirose:2004}. But, it appears that the slow wave branch is the
most strongly affected by this complication; for example, the
one-dimensional collisionless dynamics of ions along the magnetic
field leads to a parallel sound speed $C_s^2=(T_e+ 3 T_i)/m_i$
\citep{Hirose:2004}, and this produces unusual behavior, even at 
MHD scales $k \rho_i \ll 1$, such as a slow wave with a faster phase
speed than the \Alfven wave \citep{Ballai:2002}. The closest
connection between kinetic theory and the standard Hall MHD dispersion
relation appears to occur by defining an isotropic sound speed
$v_s^2=(T_e+T_i)/m_i$ \citep{Hirose:2004}, so this is the definition
taken here. Although this choice is not entirely satisfactory, we will
see that the slow wave is strongly damped via wave-particle
interactions when warm ions are present, so the differences arising
from the details of the fluid closure are often irrelevant.

The second complication in comparing Hall MHD with kinetics is that
the linear fluid modes of Hall MHD cannot be assigned a one-to-one
correspondence with the linear kinetic modes over all parameter space
\citep{Orlowski:1994,Krauss-Varban:1994,Yoon:2008}. For example, in a study 
of low-frequency waves in the solar wind upstream of Venus,
\citet{Orlowski:1994} found a correspondence to the fast Hall MHD mode
at low $\beta$ and to the \Alfven Hall MHD mode at high $\beta$, while
the results were uniformly consistent with the fast mode of kinetic
theory.  A subsequent study comparing Hall MHD with kinetic theory
found this mixing of modes to be commonplace
\citep{Krauss-Varban:1994}, perhaps accounting for the surprising 
changes in the linear wave properties of Hall MHD at small scales
noted in a recent paper \citep{Hameiri:2005}.

This mixing of solutions to the linear dispersion relations can lead
to confusion about the connections between wave modes in different
regimes of parameter space. For a chosen set of plasma parameters
$\beta_i$, $T_i/T_e$, and $v_{ti}/c$, each solution to the dispersion
relation describes a manifold in $(\mbox{Re}[\omega/\Omega_i], k_\perp
d_i, k_\parallel d_i)$ space, termed a ``dispersion surface''
\citep{Andre:1985}. A single dispersion surface may describe distinct
wave modes (with different physical properties) in different regimes
of wave vector space. For example, the
\Alfven dispersion surface describes the MHD \Alfven wave at large scale
($ k_\perp d_i \ll 1$ and $k_\parallel d_i \ll 1$), the ion cyclotron
wave at small parallel scales ($ k_\perp d_i \ll 1$ and $k_\parallel
d_i \gtrsim 1$), and the kinetic \Alfven wave at small perpendicular
scales ($ k_\perp d_i \gtrsim 1/\sqrt{\beta_i}$ and $k_\parallel d_i
\ll 1$) \citep{Yoon:2008}. In  this paper
we avoid confusion by naming the solution branch of the dispersion
relation according to the MHD mode to which it is topologically
connected along the dispersion surface.  Thus, we will investigate the
properties of the slow (S), \Alfven (A), and fast (F) branch solutions
of both the linear Hall MHD and Vlasov-Maxwell systems. Since the
solutions of these branches in the two systems do not exhibit a
one-to-one correspondence, we take the position that we want to
compare the Hall MHD wave modes with the most similar kinetic mode at
each point in parameter space, and will remark upon which modes were
chosen when relevant.

To connect quantitatively Hall MHD to kinetic theory, we must specify
the relation between $\beta_0$ in the Hall MHD dispersion relation and
$\beta_i$ in the Vlasov-Maxwell dispersion relation. For a finite ion
temperature, we choose to define $\beta_0=v_s^2/v_A^2$, where we take
an isotropic sound speed $v_s^2=(T_e+T_i)/m_i$.  Thus, the necessary
relation is given by $\beta_0=\beta_i(1+T_i/T_e)/(2 T_i/T_e)$. Note
that, in the limit $T_i\ll T_e$ in which Hall MHD is rigorously valid,
the ion pressure is ignorable and so $v_s \rightarrow
c_s=\sqrt{T_e/m_i}$, the ion acoustic speed; for turbulence in the
solar wind, the ion and electron temperatures are comparable $T_i\sim
T_e$, but standard Hall MHD has no means to account for distinct ion
and electron temperatures, so we merely quote the value of $T_i/T_e$
used for the Vlasov-Maxwell calculation.

\subsection{Comparative Measures}

Hall MHD is a fluid theory of the behavior in a magnetized plasma, and
therefore does not account for the effect of collisionless damping;
the eigenfrequencies of the Hall MHD dispersion
relation~(\ref{eq:hmhd}) are real, $\omega_h$. Therefore, in comparing
with the complex eigenfrequencies of the Vlasov-Maxwell dispersion
relation $\omega_v-i\gamma_v$, we must separately account for the
error due to the differences in the real frequencies and the error due
to the absence of damping in Hall MHD. Discrepancies in the real mode
frequencies may yield Hall MHD results for a turbulent cascade that
are quantitatively incorrect but still qualitatively correct with
respect to a kinetic solution. But without capturing the often strong
kinetic damping (via wave-particle interactions) occurring in a weakly
collisional plasma, the Hall MHD model may give results for turbulence
in the plasma that are both quantitatively and qualitatively
incorrect.

We therefore will present two measures of comparison: first, the
normalized difference between the Hall MHD frequency and the real part
of the complex kinetic frequency $\Delta \omega \equiv
|\omega_h-\omega_v|/\omega_v$; and, second, the normalized
collisionless damping rate $\gamma_v/\omega_v$ in linear kinetic
theory.  In general , a normalized damping rate of $\gamma/\omega = 1/(2 \pi)
\simeq 0.16$ is sufficient to reduce the amplitude of a wave mode by
$e^{-1}\simeq 0.37$ over a single wave period.  The position is taken
in this work that any damping rate $\gamma/\omega \ge 0.1$ is
sufficient to cause qualitative differences in the dynamics.

\section{Results}
\label{sec:results}
\subsection{Cold Ion Limit, $T_i/T_e \ll 1$}

\begin{figure}[t]
\vspace*{2mm}
\begin{center}
\includegraphics[width=8.3cm]{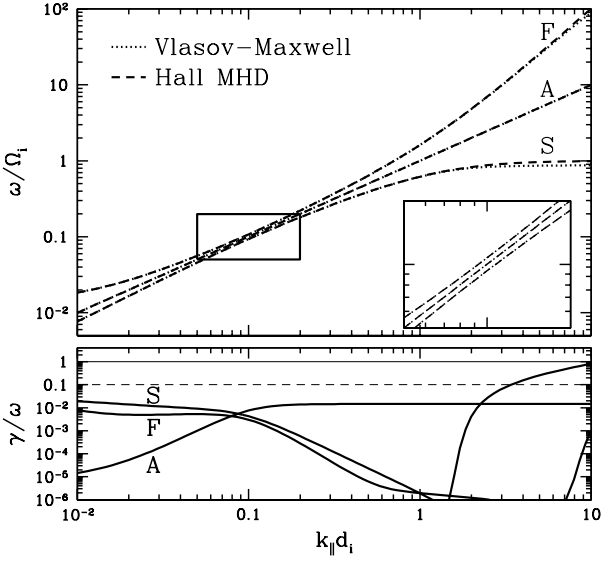}
\end{center}
\caption{Top: Normalized real frequency $\omega/\Omega_i$ vs.~parallel
wavenumber $k_\parallel d_i$ from Hall MHD (dashed) and kinetic theory
(dotted) for the fast (F), \Alfven (A), and slow (S) modes for the
parameters $\beta_0=1$, $T_i/T_e=10^{-3}$, and $k_\perp d_i
=10^{-2}$. Inset is an expanded view of the boxed region. Bottom:
Normalized damping rates $\gamma/\omega$ vs.~parallel wavenumber
$k_\parallel d_i$ for the same three low-frequency modes from kinetic
theory (solid). Values above the dashed line $\gamma/\omega=0.1$
indicate strong linear kinetic damping.}
\label{fig:kpar_cold}
\end{figure}

First, we wish to verify that standard Hall MHD and Vlasov-Maxwell
kinetic theory do agree in the limit $T_i \ll T_e$. We choose the plasma 
beta $\beta_0=1$ and take $T_i/T_e= 10^{-3}$, yielding an ion plasma
beta in the kinetic theory $\beta_i \simeq 2\times 10^{-3}$.
\figref{fig:kpar_cold} presents real frequencies 
(dashed line for Hall MHD, dotted line for Vlasov-Maxwell) in the top
panel and damping rates (solid line for Vlasov-Maxwell) in the bottom
panel for parallel wavenumbers $10^{-2} \le k_\parallel d_i \le 10$
and perpendicular wavenumber $k_\perp d_i =10^{-2}$.  The solution
labeled F is the fast magnetosonic wave at $k_\parallel d_i \ll 1$ and
the parallel whistler wave at $k_\parallel d_i \gtrsim 1$.  At
$k_\parallel d_i \ll 1$, the solution A is the shear \Alfven wave and
the solution S is the slow wave, often termed the ion acoustic wave in
this cold ion limit. In \figref{fig:kpar_cold}, the inset in the top
panel shows an expanded view of the region around $k_\parallel d_i
\sim 0.1$, demonstrating that neither the Hall MHD nor the kinetic
solutions cross. An examination of the linear kinetic eigenfunctions
(not shown) shows that the A and S solution branches undergo a mode
conversion
\citep{Swanson:1989,Stix:1992} at $k_\parallel  d_i \simeq 0.09$,
exchanging physical characteristics with each other.  Thus, at
$k_\parallel d_i \gtrsim 0.1$, the A branch corresponds to the slow
wave and the S branch corresponds to the ion cyclotron wave. This
mixing of the A and S branches occurs commonly in Hall MHD
\citep{Krauss-Varban:1994,Yoon:2008}.  In this limit $T_i \ll T_e$,
the real frequencies for all three modes are in excellent agreement;
damping is weak for all modes over this range except for the S branch
at $k_\parallel d_i > 3$, due to cyclotron damping of the
corresponding ion cyclotron waves.

\begin{figure}[t]
\vspace*{2mm}
\begin{center}
\includegraphics[width=8.3cm]{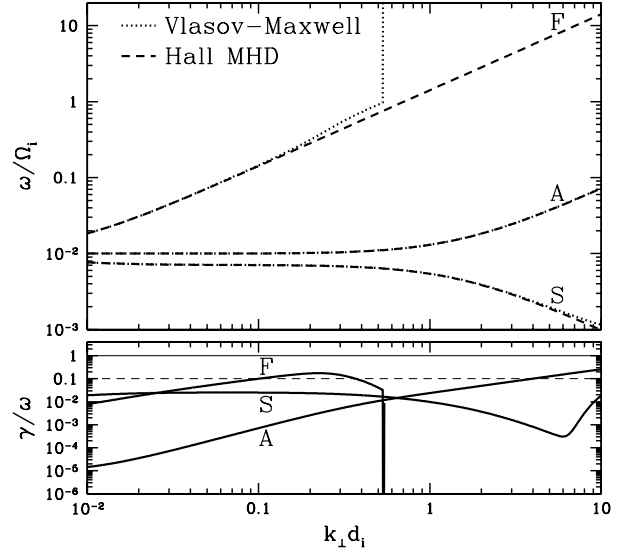}
\end{center}
\caption{Top: Normalized real frequency $\omega/\Omega_i$ vs.~perpendicular
wavenumber $k_\perp d_i$ from Hall MHD (dashed) and kinetic theory
(dotted) for the fast (F), \Alfven (A), and slow (S) modes for the
parameters $\beta_0=1$, $T_i/T_e=10^{-3}$, and $k_\parallel d_i
=10^{-2}$. Bottom: Normalized damping rates $\gamma/\omega$
vs.~perpendicular wavenumber $k_\perp d_i$ for the same three
low-frequency modes from kinetic theory (solid). }
\label{fig:kperp_cold}
\end{figure}

\figref{fig:kperp_cold} presents real frequencies and damping rates  
for perpendicular wavenumbers $10^{-2} \le k_\perp d_i \le 10$ and
parallel wavenumber $k_\parallel d_i =10^{-2}$. The Hall MHD slow wave
solution S agrees well with the weakly damped kinetic solution.  The
\Alfven wave mode A also shows good agreement, but after it
transitions to the kinetic \Alfven wave at $k_\perp d_i \sim 1$,
damping via the Landau resonance becomes significant at $k_\perp d_i >
4$. The fast wave solution F from kinetic theory, on the other hand,
deviates from the Hall MHD solution at $k_\perp d_i > 0.5$; here the
numerical solver used is unable to follow the fast wave root as it
undergoes a mode conversion into an ion Bernstein mode. Bernstein
waves arise due to finite Larmor radius effects that have no
counterpart in cold plasma theory and that are not described by Hall
MHD \citep{Swanson:1989}; the impact of the ion Bernstein waves is
discussed at length in \secref{sec:ibw}.

Despite some of these difficulties with ion Bernstein waves and
non-negligible kinetic damping, Figures~\ref{fig:kpar_cold} and
\ref{fig:kperp_cold} bear out the general finding
\citep{Hirose:2004,Ito:2004} that Hall MHD is a valid limit of kinetic theory 
in the cold ion approximation, $T_i \ll T_e$.

\subsection{Finite Ion Temperature $T_i$}
\label{sec:fti}
\begin{figure}[t]
\vspace*{2mm}
\begin{center}
\includegraphics[width=8.3cm]{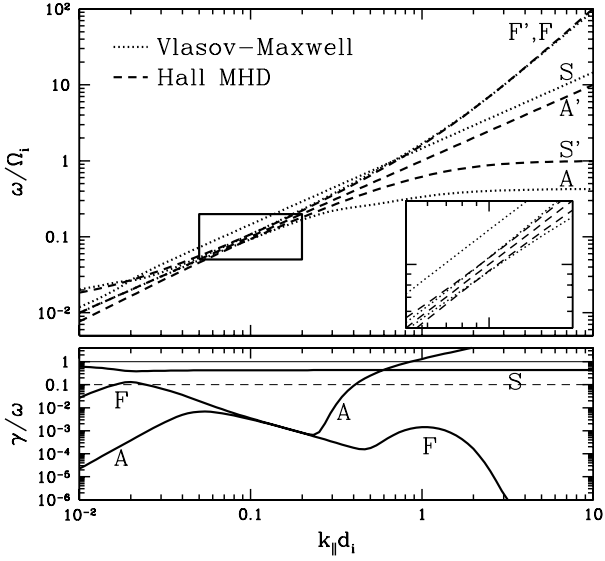}
\end{center}
\caption{Top: Normalized real frequency $\omega/\Omega_i$ vs.~parallel
wavenumber $k_\parallel d_i$ for the Hall MHD (dashed) fast (F'),
\Alfven (A'), and slow (S') modes and for the kinetic theory (dotted)
fast (F), \Alfven (A), and slow (S) modes for the parameters
$\beta_0=1$, $T_i/T_e=1$, and $k_\perp d_i =10^{-2}$.  Inset is an
expanded view of the boxed region.  Bottom: Normalized damping rates
$\gamma/\omega$ vs.~parallel wavenumber $k_\parallel d_i$ for the same
three low-frequency modes from kinetic theory (solid).}
\label{fig:kpar_warm}
\end{figure}

\begin{figure}[t]
\vspace*{2mm}
\begin{center}
\includegraphics[width=8.3cm]{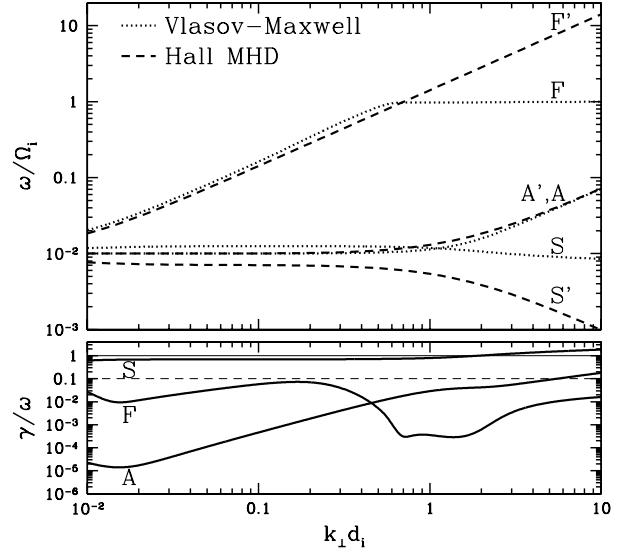}
\end{center}
\caption{Top: Normalized real frequency $\omega/\Omega_i$ vs.~perpendicular
wavenumber $k_\perp d_i$  for the Hall MHD (dashed) fast (F'),
\Alfven (A'), and slow (S') modes and for the kinetic theory (dotted)
fast (F), \Alfven (A), and slow (S) modes for the parameters
$\beta_0=1$, $T_i/T_e=1$, and $k_\parallel d_i
=10^{-2}$. Bottom: Normalized damping rates $\gamma/\omega$
vs.~perpendicular wavenumber $k_\perp d_i$ for the same three
low-frequency modes from kinetic theory (solid).}
\label{fig:kperp_warm}
\end{figure}

Next, we explore the limit of warm ions relevant to most space and
astrophysical plasmas.  We take $\beta_0=1$ and $T_i/T_e= 1$, giving
an ion plasma beta in the kinetic theory $\beta_i = 1$.
\figref{fig:kpar_warm} presents real frequencies (top) and damping rates (bottom)
for parallel wavenumbers $10^{-2} \le k_\parallel d_i \le 10$ and
perpendicular wavenumber $k_\perp d_i =10^{-2}$. The frequency of the
kinetic fast wave F (dotted) is well reproduced by Hall MHD solution
F' (dashed), and its damping is weak except for a small region around
$k_\parallel d_i \sim 2 \times 10^{-2}$. Comparison of the slow and
\Alfven wave solutions demonstrates some of the complications arising in a
comparison of Hall MHD to kinetic theory as described in
\secref{sec:subtle}.  The kinetic slow wave solution S (dotted) has a phase
speed that is faster than that of the \Alfven wave solution A (dotted)
of kinetic theory (this is perhaps more easily seen
\figref{fig:kperp_warm}); the anisotropic ion pressure response in the
kinetic theory that leads to this curious result \citep{Ballai:2002}
is not captured by the isotropic equation of state used in the
standard Hall MHD model. But this error is overshadowed by a far more
serious discrepancy between Hall MHD and kinetic theory: at finite ion
temperature the kinetic slow wave S (dotted)---the ion acoustic
wave---is heavily damped with $\gamma_v/\omega_v
\sim 1$, yet a fluid theory such as Hall MHD does not account for this 
damping. Therefore, at finite ion temperature the Hall MHD slow wave
represents an unphysical, spurious wave that does not exist in a weakly
collisional plasma.

The second complication apparent in \figref{fig:kpar_warm} is the
mixing of the modes between the solution branches of kinetic theory
and those of Hall MHD \citep{Krauss-Varban:1994,Yoon:2008}.  The inset
in the top panel of \figref{fig:kpar_warm} shows that the Hall MHD
solutions do not cross, however the correspondence between kinetic A
and S branches and Hall MHD A' and S' branches changes with increasing
$k_\parallel d_i$: at $k_\parallel d_i \ll 1$, the solution A (dotted)
of kinetic theory agrees closely with the Hall MHD solution A'
(dashed); but at $k_\parallel d_i > 0.1$, the solution A (dotted) from
kinetic theory (the ion cyclotron wave) appears to correspond to the
Hall MHD solution S' (dashed) and the kinetic solution S (dotted)
appears to correspond most closely with the Hall MHD solution A'.
This mixing up of fluid and kinetic modes complicates any attempted
comparison of the two models, but a careful matching of the most
similar wave modes will yield a fair evaluation. Again, however, the
major discrepancy between the models is that Hall MHD does not capture
the strong kinetic damping of the kinetic S branch at all scales and
of the solution A from kinetic theory (the ion cyclotron wave) at
$k_\parallel d_i > 0.4$ .

\figref{fig:kperp_warm} presents real frequencies and damping rates  
for perpendicular wavenumbers $10^{-2} \le k_\perp d_i \le 10$ and
parallel wavenumber $k_\parallel d_i =10^{-2}$ for the warm ion case
with $\beta_0=1$ and $T_i/T_e= 1$.  The kinetic branch S is once again
strongly damped by the warm ions, so Hall MHD supports a spurious,
undamped slow wave mode S'.  The kinetic branch A, representing MHD
\Alfven waves and kinetic \Alfven waves, is quite well represented
over the entire wavenumber range, with some discrepancy near $k_\perp
d_i \sim 1$. The kinetic fast wave solution F is well matched by the
Hall MHD fast wave solution F', with only a slight deviation, for
$k_\perp d_i \lesssim 0.4$. At $k_\perp d_i = 0.4$, however, the kinetic branch
F goes through a mode conversion to become an ion Bernstein wave, a
situation discussed at length in
\secref{sec:ibw}.

\begin{figure}[t]
\vspace*{2mm}
\begin{center}
\includegraphics[width=8.3cm]{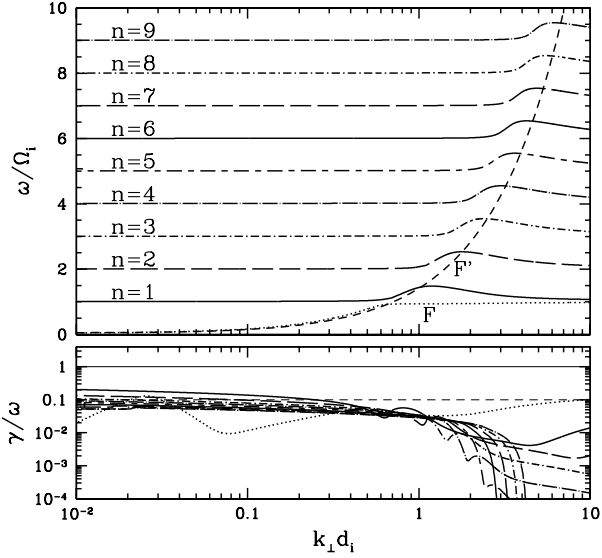}
\end{center}
\caption{Top: Normalized real frequency $\omega/\Omega_i$ vs.~perpendicular
wavenumber $k_\perp d_i$ for the Hall MHD fast mode (F', dashed), the
kinetic theory fast mode (F, dotted), and the kinetic $n=1$ to $n=9$
ion Bernstein wave modes for the parameters $\beta_0=1$, $T_i/T_e=1$,
and $k_\parallel d_i =5 \times 10^{-2}$. Bottom: Normalized damping
rates $\gamma/\omega$ vs.~perpendicular wavenumber $k_\perp d_i$ for
the modes from kinetic theory, the fast (dotted) and $n=1$ to $n=9$
ion Bernstein wave modes.}
\label{fig:ibw}
\end{figure}

In summary, at finite ion temperature, Hall MHD does a good job
modeling the kinetic fast wave branch F for nearly parallel wave
vectors and the kinetic \Alfven wave branch for nearly perpendicular
wave vectors. But, it fails to capture the strong kinetic damping of
the slow mode branch in all cases and of the \Alfven branch in the
nearly parallel case. The correspondence of the Hall MHD fast wave
branch F' at nearly perpendicular wave vectors to a series of kinetic
ion Bernstein waves is discussed next.

\subsection{Ion Bernstein Waves}
\label{sec:ibw}

The fast solution F from kinetic theory in \figref{fig:kperp_warm}
deviates greatly from the Hall MHD fast solution F' at $k_\perp d_i
\gtrsim 0.4$. The cause of this deviation is the conversion, in 
kinetic theory, from an MHD fast wave to an $n=1$ ion Bernstein wave
(with $\omega < \Omega_i$) on the dispersion surface of solution
branch F. Not shown in \figref{fig:kperp_warm} is the conversion of an
$n=1$ ion Bernstein wave (with $\omega > \Omega_i$) to a continuation
of the fast wave at $\omega > \Omega_i$; this wave lies on a different
dispersion surface from branch F. Mode conversion from fast wave
solutions to ion Bernstein waves at nearly perpendicular propagation
is a well known physical phenomenon
\citep{Swanson:1989,Stix:1992,LiHabbal:2001}.  Here we aim  to 
evaluate the accuracy of the Hall MHD description of the kinetic
plasma behavior in this regime.

In \figref{fig:ibw} are presented the real frequencies (top panel) and
damping rates (bottom panel) for the Hall MHD fast branch F' (dashed),
the kinetic fast branch F (dotted), and the kinetic $n=1$ through
$n=9$ ion Bernstein solutions for perpendicular wavenumbers $10^{-2}
\le k_\perp d_i \le 10$ and parallel wavenumber $k_\parallel d_i =5
\times 10^{-2}$ for $\beta_0=1$ and $T_i/T_e= 1$. Note that each of the 
ion Bernstein solutions lie on distinct dispersion surfaces.  In the
top panel, the relation between the Hall MHD branch F' and the series
of ion Bernstein wave solutions is clear, yet the summed effect of
these kinetic wave modes differs significantly in a number of ways
from the Hall MHD fast branch solution F'. First, between each of the
ion Bernstein solutions lies a frequency gap in which no solution
exists. Second, the real frequency of the ion Bernstein waves (and
thus the perpendicular phase velocity $\omega/k_\perp$) is only
coincident with the frequency given by Hall MHD at a single point for
each of the the $n \ge 2$ Bernstein modes. Finally, although the
perpendicular group velocity (the slope of the plot $\partial
\omega/\partial k_\perp$) of the Hall MHD F' solution is always
positive and increasing with $k_\perp d_i$, each of the ion Bernstein
waves have a positive but smaller group velocity at lower $k_\perp
d_i$, and a negative group velocity at higher $k_\perp d_i$,
corresponding to a backward wave \citep{Stix:1992} at these scales.
The lower panel of \figref{fig:ibw} shows that the kinetic fast branch
F and $n=1$ through $n=9$ ion Bernstein waves are not strongly damped.
Our conclusion is that the kinetic dynamics of the fast branch for
nearly perpendicular wave vectors in a weakly collisional plasma is
not well represented by Hall MHD.

\subsection{Comparison on the $(k_\perp,k_\parallel)$ Plane}
\label{sec:kpkz}

The axisymmetric nature of any uniform magnetized plasma about the
direction of a uniform, straight magnetic field $\V{B}_0$ reduces the
dependence on the wave vector $\V{k}$ to its two components
perpendicular and parallel to the magnetic field,
$(k_\perp,k_\parallel)$; this is clearly seen in the dispersion
relations of both the Hall MHD and the Vlasov-Maxwell systems with
their dependence on the parameters $k_\perp d_i$ and $k_\parallel
d_i$. Therefore, plots of the real frequency difference $\Delta
\omega$ and of the kinetic damping rate $\gamma_v/\omega_v$ on the plane
$(k_\perp,k_\parallel)$ provide excellent quantitative measures of the
fidelity of Hall MHD to the kinetic physics in a weakly collisional
plasma over a wide range of scales. 

\begin{figure*}[t]
\vspace*{2mm}
\begin{center}
\hbox to \hsize{\resizebox{8.3cm}{!}{
\includegraphics*[0.6in,2.75in][7.3in,7.95in]{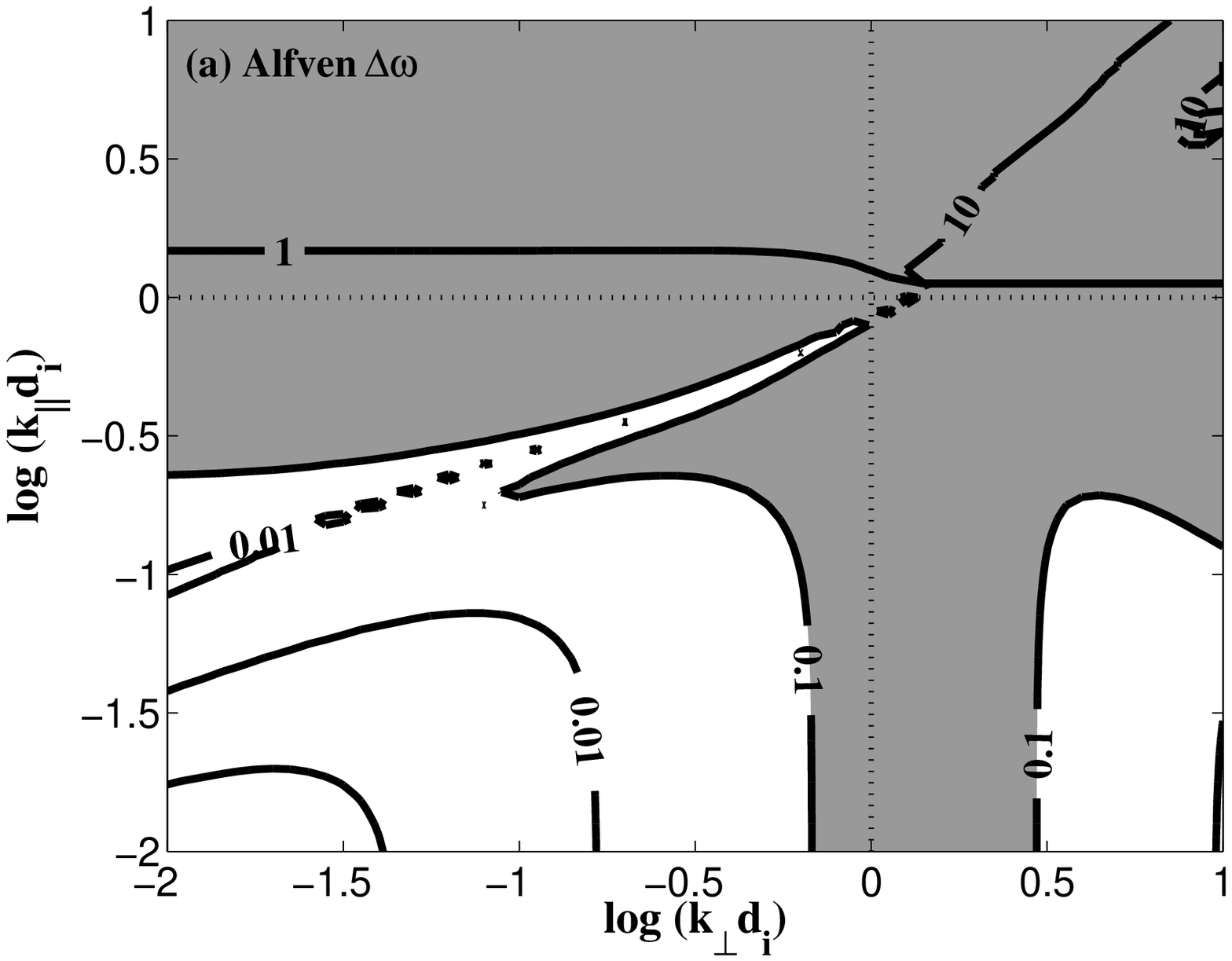}} \hfill
\resizebox{8.3cm}{!}{
\includegraphics*[0.6in,2.75in][7.3in,7.95in]{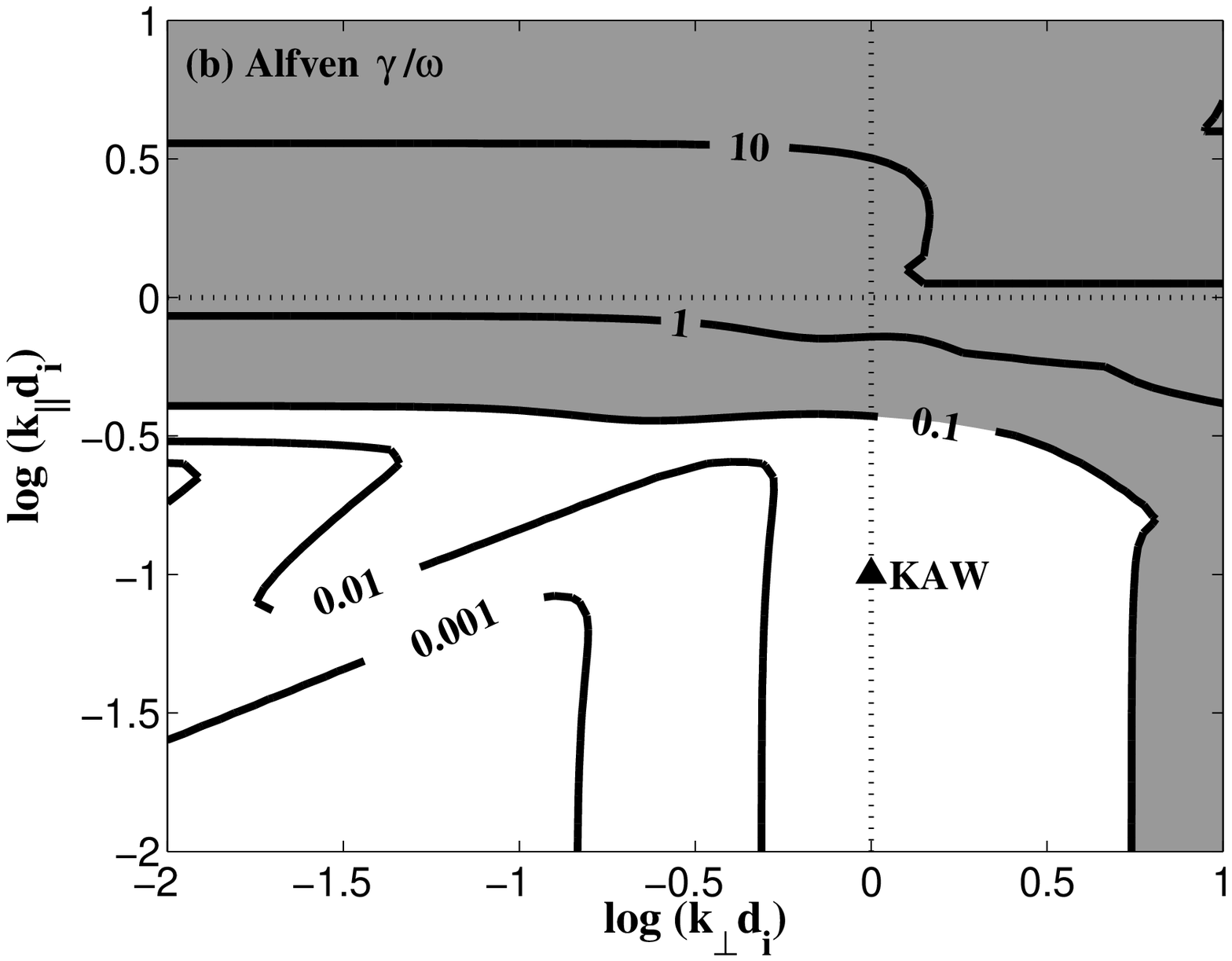}}}
\end{center}
\caption{(a)~Normalized difference in the real frequency 
$\Delta \omega \equiv |\omega_h -\omega_v| /\omega_v$ for the \Alfven
wave branch on the $(k_\perp,k_\parallel)$ plane for $\beta_i=1$ and
$T_i/T_e=1$. Shading denotes values of $\Delta \omega >1$,
corresponding to a parameter regime where Hall MHD differs
significantly from Vlasov-Maxwell kinetic theory. (b)~The
corresponding normalized linear damping rate $\gamma_v/\omega_v$,
where shading of areas with $\gamma_v/\omega_v>1$ highlights regimes
in which the collisionless damping of kinetic theory is substantial.}
\label{fig:kpkz_alf}
\end{figure*}

In panel (a) of \figref{fig:kpkz_alf}, the normalized real frequency
difference between Hall MHD and Vlasov-Maxwell kinetics, $\Delta
\omega \equiv |\omega_h-\omega_v|/\omega_v$, is plotted for the
\Alfven solution over $10^{-2} \le k_\perp d_i \le 10$ and $10^{-2}
\le k_\parallel d_i \le 10$; in panel (b) is plotted the normalized
linear kinetic damping rate, $\gamma_v/\omega_v$. Plasma parameters
relevant to the near-earth solar wind, $\beta_i=1$ and $T_i/T_e=1$,
are chosen. Areas of panel (a) with $\Delta \omega >0.1$ are shaded to
highlight the regions of parameter space over which the frequency
using Hall MHD deviates significantly from the kinetic value; as
evident, the \Alfven solution from Hall MHD\footnote{Because of the
tendency for the kinetic modes to correspond to different fluid modes
as parameters vary, as discussed in \secref{sec:subtle}, in this paper
the calculation of $\Delta \omega$ at each point uses the Hall MHD
mode that the yields the minimum value of this difference.} deviates
substantially from the Vlasov-Maxwell frequency at $k_\parallel d_i
\gtrsim 1$ and moderately at $k_\perp d_i \sim 1$. Areas of panel (b) 
with $\gamma_v/\omega_v>0.1$ are shaded to denote regions of parameter
space over which the linear collisionless damping of this mode in
Vlasov-Maxwell kinetic theory is strong. It is evident that for
$k_\parallel d_i \gtrsim 1$, the regime corresponding to ion cyclotron
waves, strong damping occurs via the ion cyclotron resonance; it seems
likely that this strong damping, due to  wave-particle interactions not
captured by Hall MHD, is also responsible for the poor agreement in
real frequency. Also evident in panel (b) is the increasingly strong
damping of the \Alfven solution at $k_\perp d_i \gg 1$ and $k_\perp
\gg k_\parallel$; this is the regime of kinetic \Alfven waves in which
damping occurs via Landau resonance with the electrons.

\begin{figure*}[t]
\vspace*{2mm}
\begin{center}
\hbox to \hsize{\resizebox{8.3cm}{!}{
\includegraphics*[0.6in,2.75in][7.3in,7.95in]{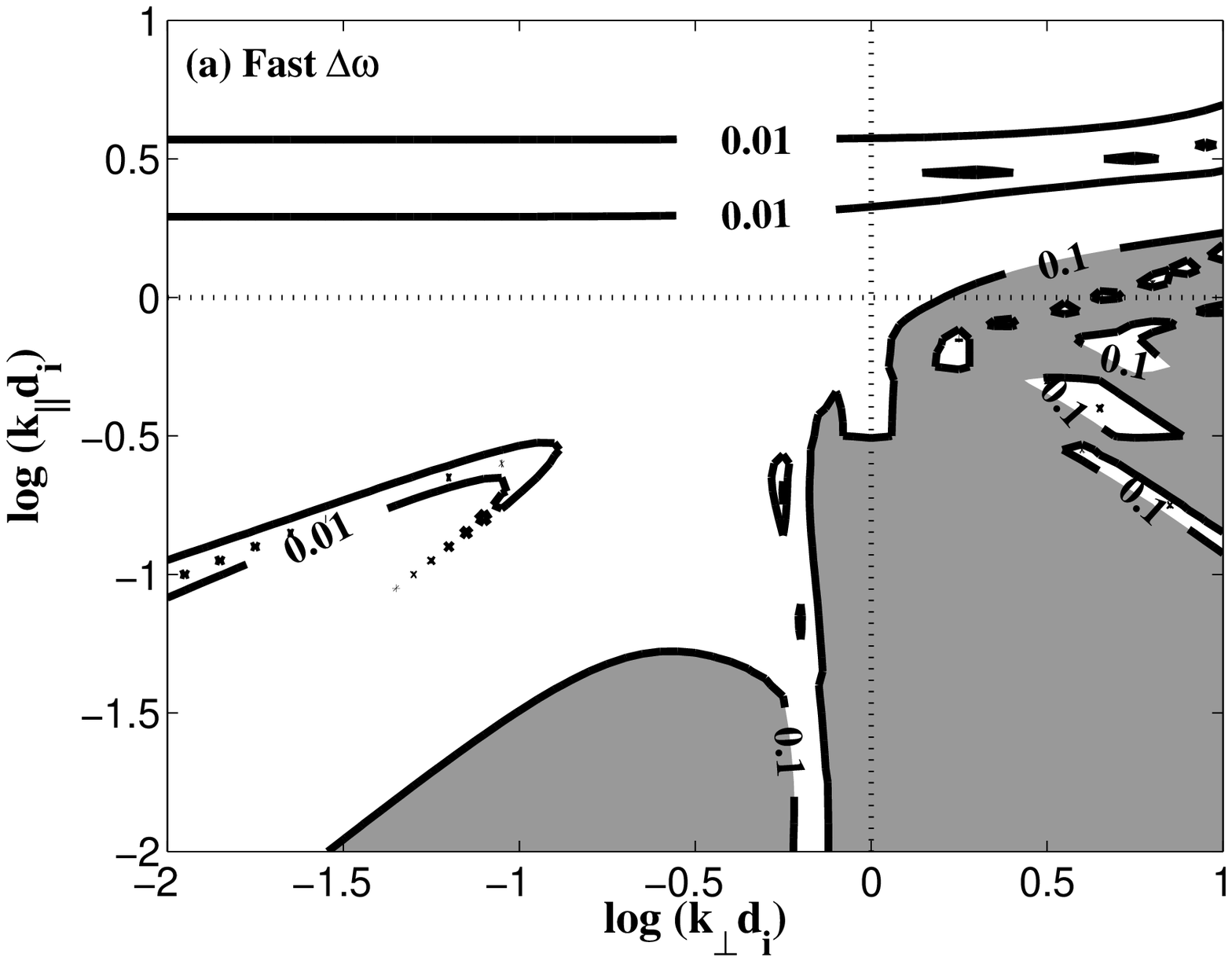}} \hfill
\resizebox{8.3cm}{!}{
\includegraphics*[0.6in,2.75in][7.3in,7.95in]{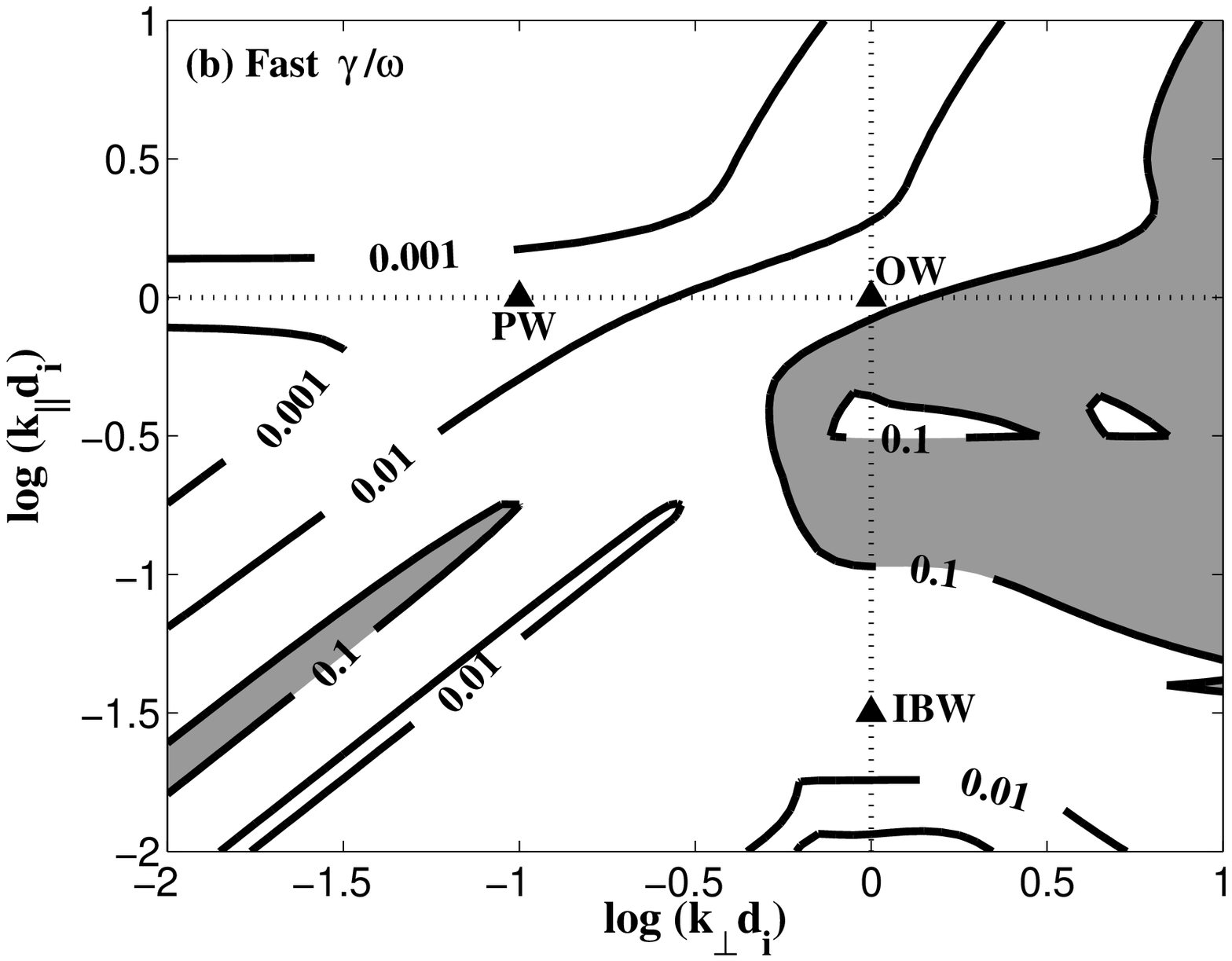}}}
\end{center}
\caption{(a)~Normalized difference in the real frequency 
$\Delta \omega$ for the fast wave branch on the
$(k_\perp,k_\parallel)$ plane for $\beta_i=1$ and $T_i/T_e=1$. (b)~The
corresponding normalized linear damping rate $\gamma_v/\omega_v$.}
\label{fig:kpkz_fas}
\end{figure*}

The analogous calculation for the fast wave branch is presented in 
\figref{fig:kpkz_fas}, with the frequency difference $\Delta \omega$
plotted in panel (a) and the kinetic damping rate $\gamma_v/\omega_v$
in panel (b), again for plasma parameters $\beta_i=1$ and $T_i/T_e=1$.
Note that this calculation employs only the kinetic solutions on the
dispersion surface of the F branch and does not consider the $n\ge1$
ion Bernstein wave dispersion surfaces.  The real frequency derived
from Hall MHD differs from the kinetic value in two regimes: a
moderate difference occurs at $k_\parallel d_i\ll 1$ and $0.05
\lesssim k_\perp d_i \lesssim 0.5$; and the difference becomes more
dramatic at $k_\perp d_i \gtrsim 1$ and $k_\parallel d_i\lesssim 1$
due to the mode conversion in kinetic theory of the F branch from the
fast wave to an ion Bernstein wave in this regime. The kinetic
damping, shown in panel (b), is relatively weak except for the regime
of the ion Bernstein waves, $k_\perp d_i \gtrsim 1$ and $k_\parallel
d_i\lesssim 1$; these waves are strongly damped unless $k_\parallel
d_i\ll 1$ \citep{Swanson:1989}.

The slow wave branch suffers very strong linear collisionless damping
in a plasma with finite ion temperature, such as the solar wind, over
the entire range of scales considered in
Figures~\ref{fig:kpkz_alf}~and~\ref{fig:kpkz_fas}. A comparison of
the real frequency in this case would be meaningless since a wave with
a damping rate $\gamma/\omega \sim 1$ is so heavily damped that it can
scarcely be considered a wave.

\subsection{Plasma Parameter $\beta_i$ and  $T_i/T_e$ Variations}
\label{sec:bt}
A very important aspect of the behavior of weakly collisional plasmas
is made clear by the damping rate plots in panel (b) of both
Figures~\ref{fig:kpkz_alf}~and~\ref{fig:kpkz_fas}: at small length
scales, with wavenumbers $k d_i \gtrsim 1$, there exist only four
regimes in which undamped wave modes occur.  These are:
\begin{enumerate}
\item Kinetic \Alfven Waves: The \Alfven branch solution of 
Vlasov-Maxwell kinetic theory in the nearly perpendicular regime
$k_\parallel d_i\ll 1$ and $k_\perp d_i \gtrsim 1$ corresponds to the
kinetic \Alfven wave mode (KAW) .
\item Parallel Whistler Waves: The fast  branch solution in 
the regime $k_\parallel d_i \gtrsim 1$ and $k_\perp d_i\ll 1$
corresponds to the whistler wave with a nearly parallel wave vector (PW).
\item Oblique Whistler Waves: The fast  branch solution in 
the regime $k_\parallel d_i \gtrsim 1$ and $k_\perp d_i\gtrsim 1$
corresponds to the whistler wave with an oblique wave vector (OW).
\item Ion Bernstein Waves: The fast  branch solution in 
the nearly perpendicular regime $k_\parallel d_i \ll 1$ and $k_\perp
d_i\gtrsim 1$ corresponds to an ion Bernstein wave (IBW).
\end{enumerate}
Since the plasma turbulence at small length scales---for example,  the
dissipation range in the solar wind---is likely to consist of one, or a
mixture, of these four undamped modes, we would like to explore the
ability of Hall MHD to reproduce their linear wave frequencies
accurately (relative to linear Vlasov-Maxwell kinetic theory).  We
note, however, that the ion Bernstein wave is an electrostatic wave
mode \citep{Swanson:1989,Stix:1992}, and therefore has no
corresponding magnetic field fluctuation; since the solar wind
dissipation range is observed to have magnetic fluctuations 
\citep{Goldstein:1995,Leamon:1998a,Smith:2006,Hamilton:2008},
we do not further pursue the ion Bernstein waves here, but rather we
focus solely on the first three modes, all of which are
electromagnetic.

\begin{figure*}[t]
\vspace*{2mm}
\begin{center}
\hbox to \hsize{\resizebox{8.3cm}{!}{
\includegraphics*[0.6in,2.75in][7.3in,7.95in]{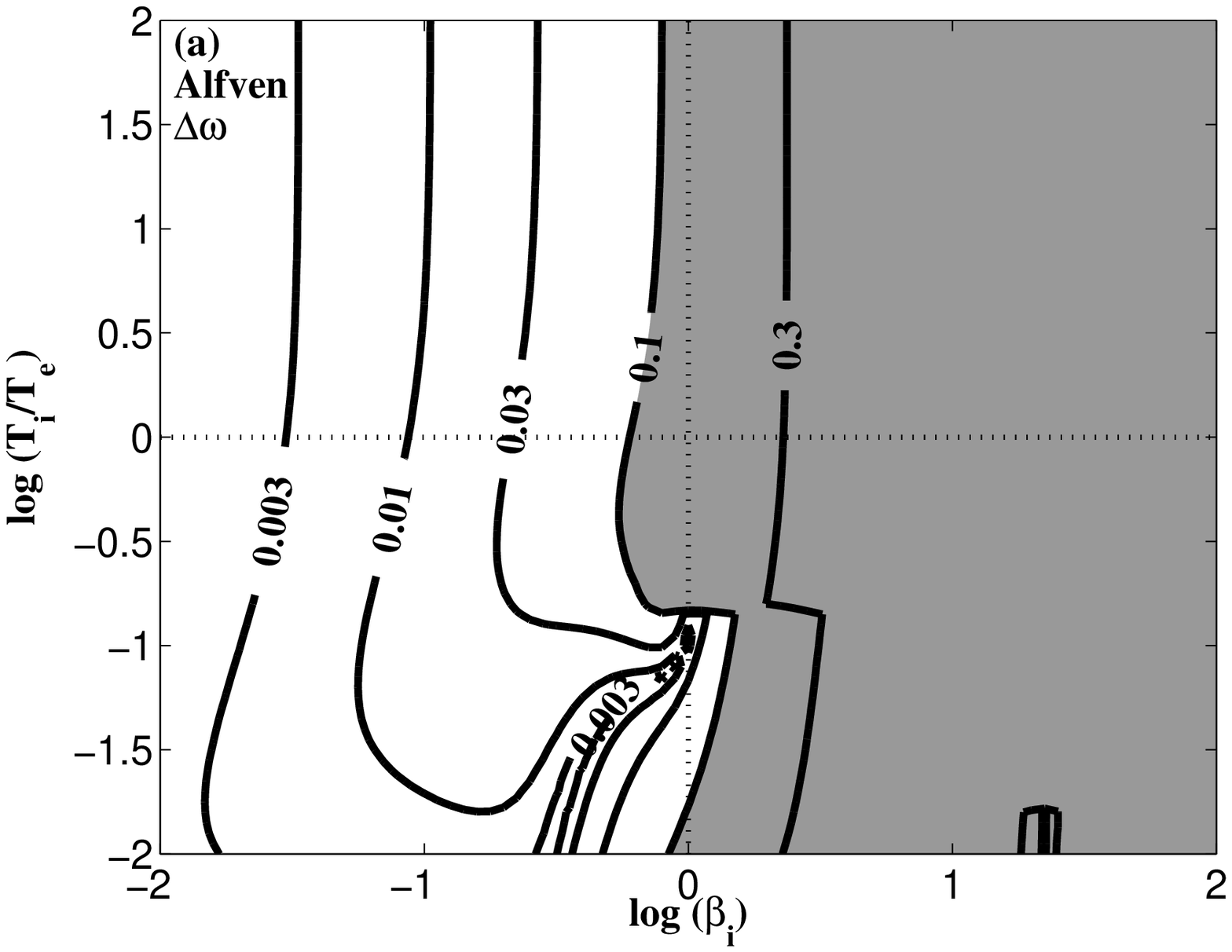}} \hfill
\resizebox{8.3cm}{!}{
\includegraphics*[0.6in,2.75in][7.3in,7.95in]{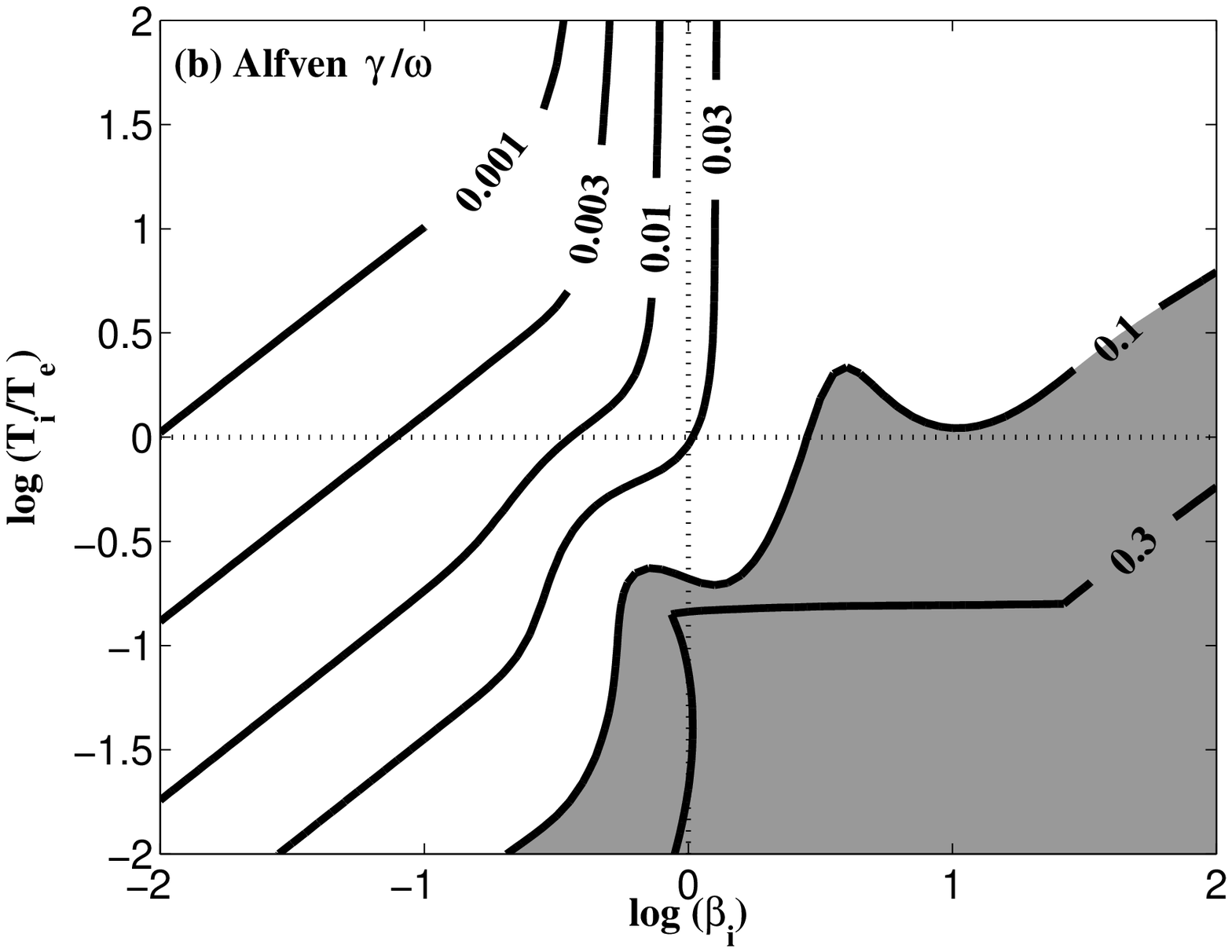}}}
\end{center}
\caption{(a)~Normalized difference in the real frequency 
$\Delta \omega$ for the kinetic \Alfven wave at $k_\perp d_i=1$ and
$k_\parallel d_i=0.1$ on the plane $(\beta_i,T_i/T_e)$. (b)~The
corresponding normalized linear damping rate $\gamma_v/\omega_v$.}
\label{fig:bt_kaw}
\end{figure*}

For a kinetic \Alfven wave at $k_\perp d_i=1$ and $k_\parallel
d_i=0.1$, denoted by the triangle (KAW) in panel (b) of
\figref{fig:kpkz_alf}, we plot the frequency difference $\Delta \omega$
and kinetic damping rate $\gamma_v/\omega_v$ over $0.01 \le \beta_i
\le 100$ and $0.01 \le T_i/T_e\le 100$  in panels (a) and (b) 
of \figref{fig:bt_kaw}. The kinetic \Alfven wave appears to be well
reproduced by Hall MHD for  $ \beta_i < 1$, but is less accurately determined
for $ \beta_i \gtrsim 1$. The linear kinetic damping of the mode 
appears to become strong only for  $ \beta_i \gtrsim 1$ and 
$ T_i/T_e \lesssim 1$.

\begin{figure*}[t]
\vspace*{2mm}
\begin{center}
\hbox to \hsize{\resizebox{8.3cm}{!}{
\includegraphics*[0.6in,2.75in][7.3in,7.95in]{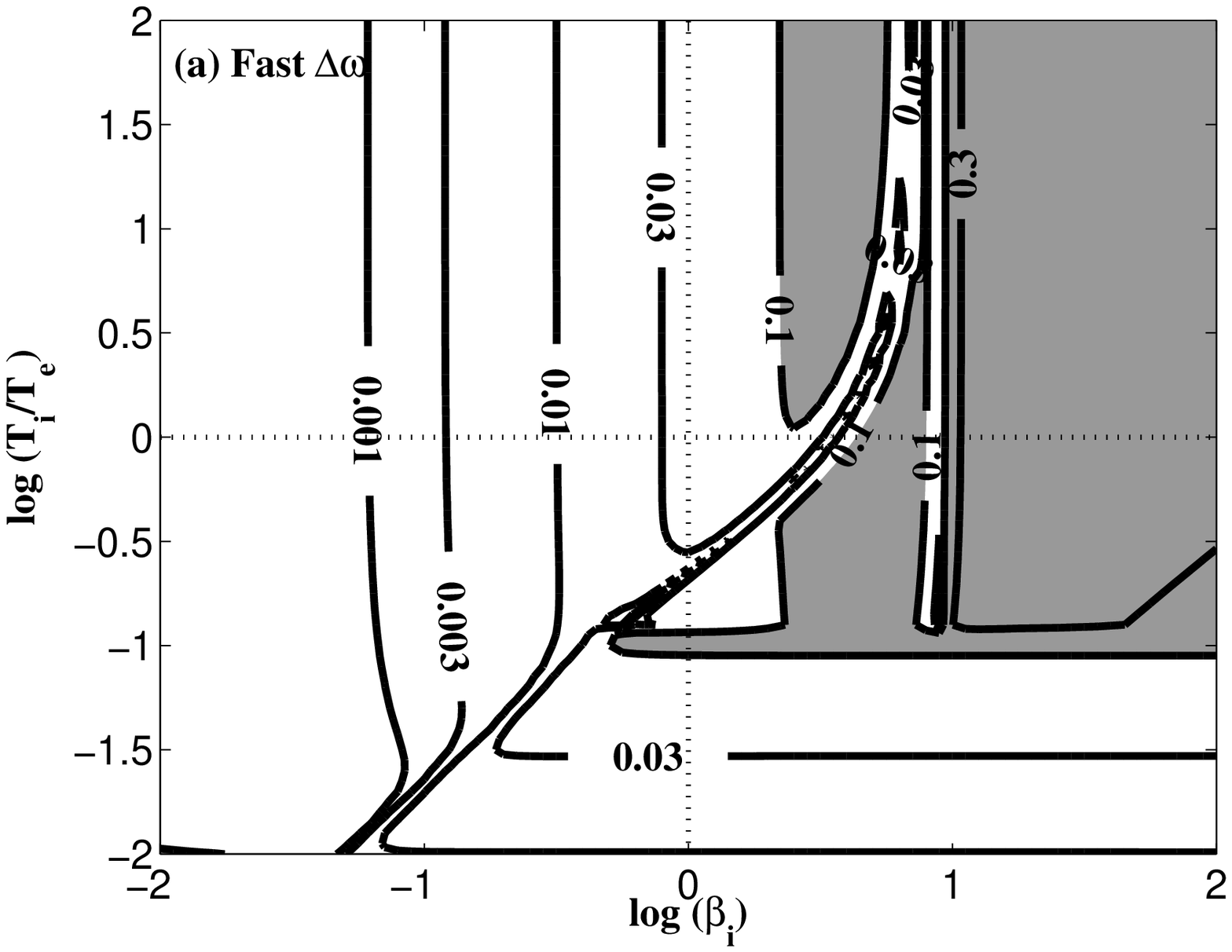}} \hfill
\resizebox{8.3cm}{!}{
\includegraphics*[0.6in,2.75in][7.3in,7.95in]{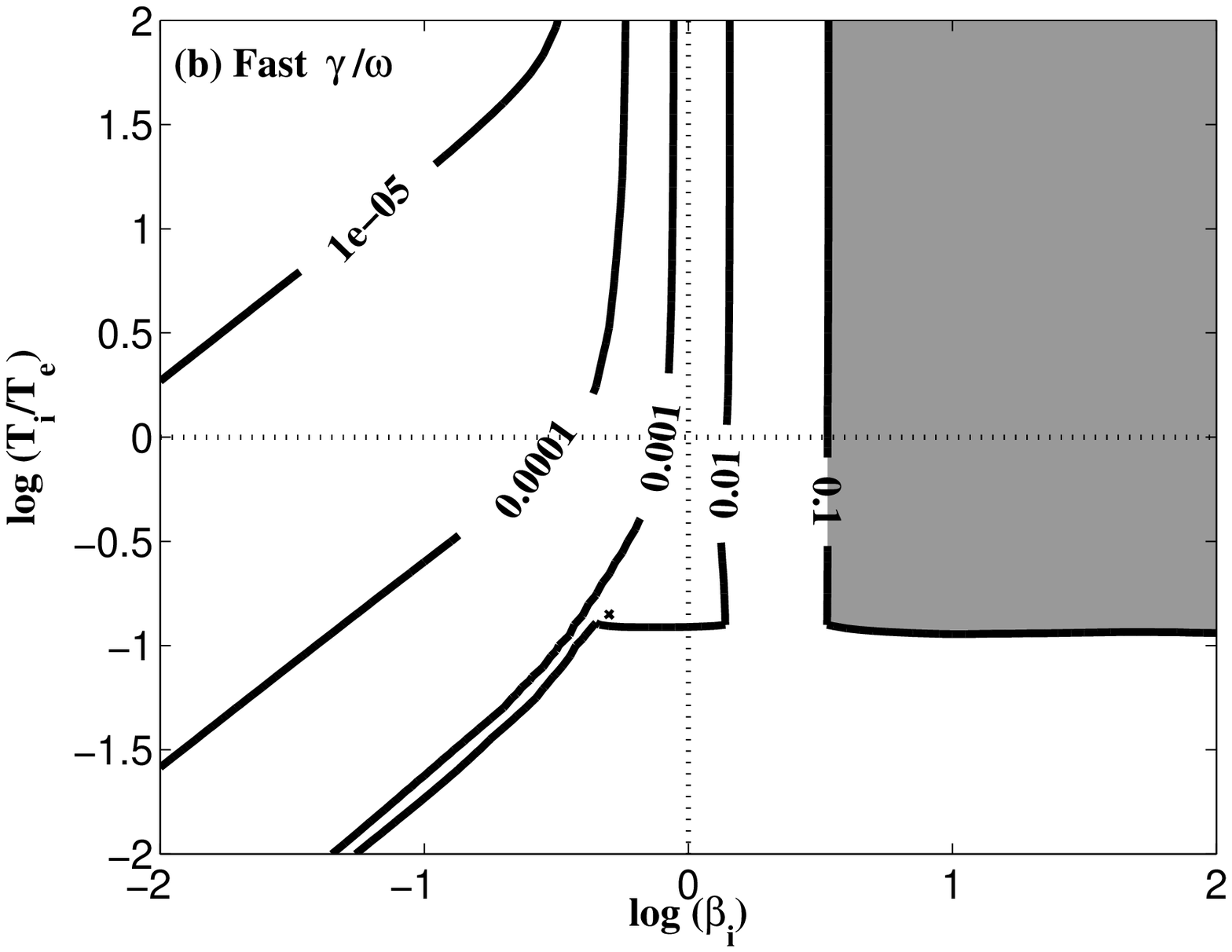}}}
\end{center}
\caption{(a)~Normalized difference in the real frequency 
$\Delta \omega$ for the parallel whistler wave at $k_\perp d_i=0.1$ and
$k_\parallel d_i=1$ on the plane $(\beta_i,T_i/T_e)$. (b)~The
corresponding normalized linear damping rate $\gamma_v/\omega_v$.}
\label{fig:bt_pw}
\end{figure*}

For a parallel whistler wave (on the fast wave branch) at $k_\perp
d_i=0.1$ and $k_\parallel d_i=1$, denoted by the triangle (PW) in
panel (b) of \figref{fig:kpkz_fas}, we plot the frequency difference
and kinetic damping rate on the $(\beta_i,T_i/T_e)$ plane in panels
(a) and (b) of \figref{fig:bt_pw}. This parallel whistler wave appears
to be well reproduced by Hall MHD over all parameter space
$(\beta_i,T_i/T_e)$ except for $ \beta_i \gtrsim 3$ and $ T_i/T_e
\gtrsim 0.1$; in this problematic regime, the whistler wave mode
converts to an ion Bernstein wave and is therefore both strongly
damped and poorly described by Hall MHD.

\begin{figure*}[t]
\vspace*{2mm}
\begin{center}
\hbox to \hsize{\resizebox{8.3cm}{!}{
\includegraphics*[0.6in,2.75in][7.3in,7.95in]{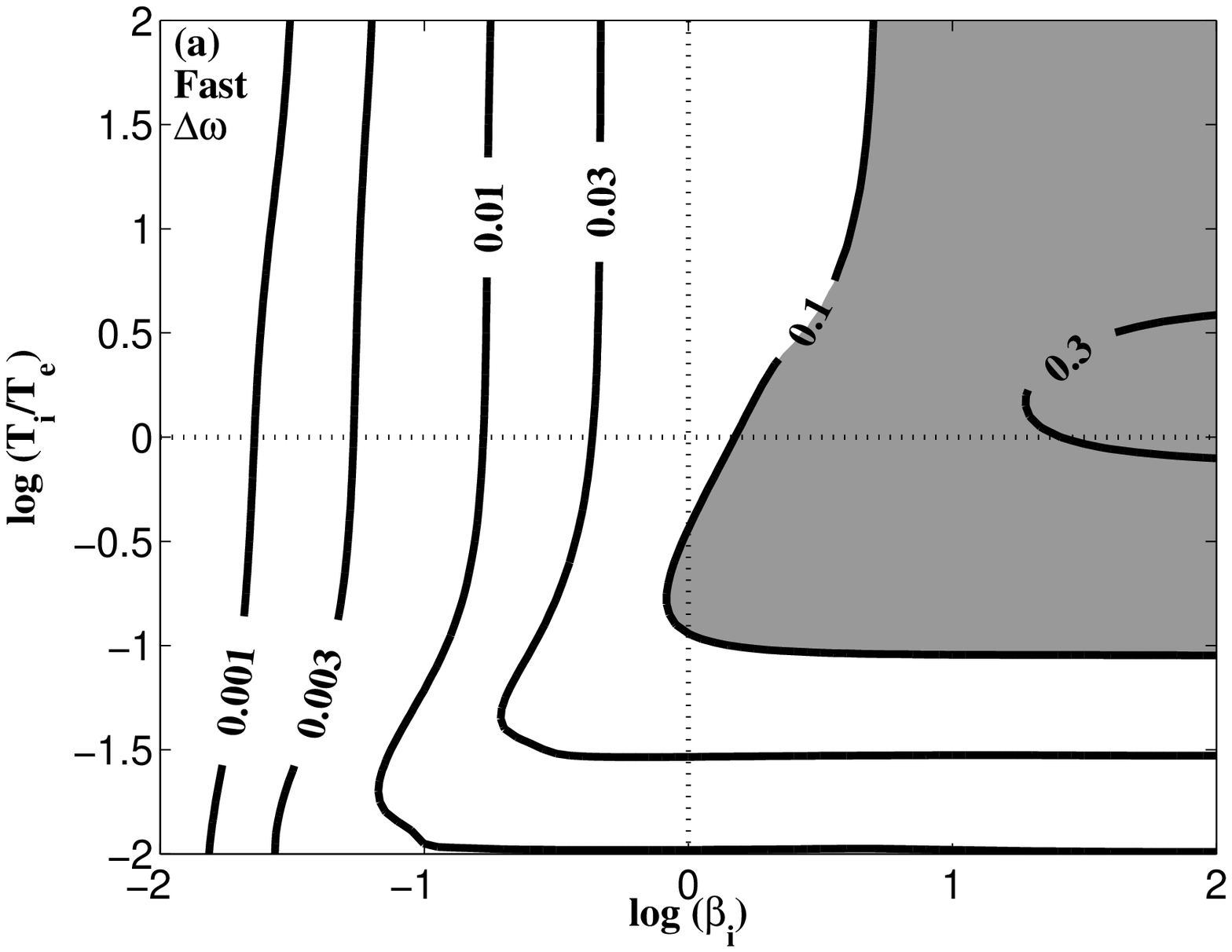}} \hfill
\resizebox{8.3cm}{!}{
\includegraphics*[0.6in,2.75in][7.3in,7.95in]{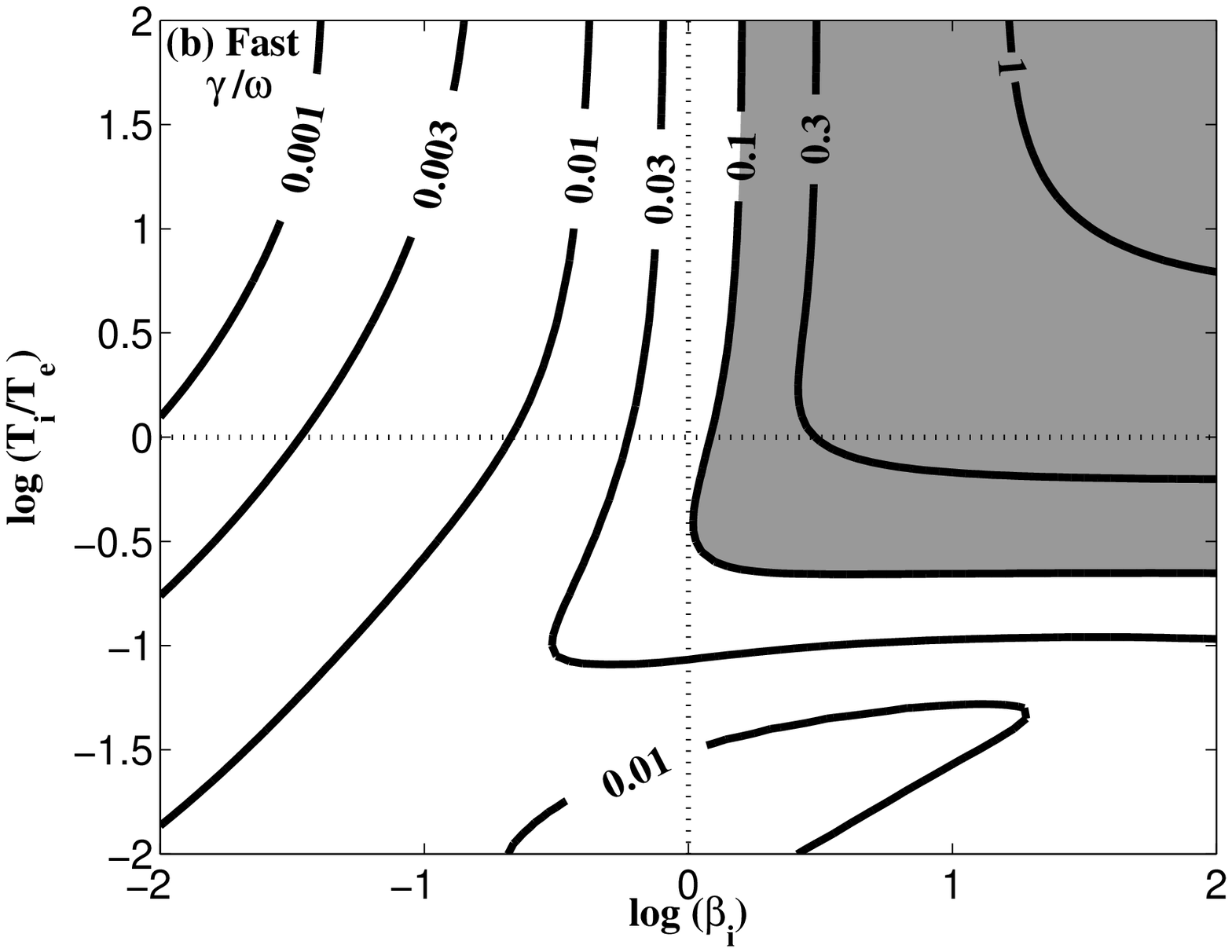}}}
\end{center}
\caption{(a)~Normalized difference in the real frequency 
$\Delta \omega$ for the oblique whistler wave at $k_\perp d_i=1$ and
$k_\parallel d_i=1$ on the plane $(\beta_i,T_i/T_e)$. (b)~The
corresponding normalized linear damping rate $\gamma_v/\omega_v$.}
\label{fig:bt_ow}
\end{figure*}

For an oblique whistler wave (on the fast wave branch) at $k_\perp d_i=1$
and $k_\parallel d_i=1$, denoted by the triangle (OW) in panel (b) of
\figref{fig:kpkz_fas}, we plot the frequency difference and kinetic
damping rate on the $(\beta_i,T_i/T_e)$ plane in
\figref{fig:bt_ow}. The results are very similar  to those for the
parallel whistler wave, with strong damping, and correspondingly poor
 agreement in linear real frequency, occurring over the regime $ \beta_i
 \gtrsim 1$ and $ T_i/T_e \gtrsim 0.1$.

With the exception of the regime $ \beta_i \gtrsim 1$ and $ T_i/T_e
\gtrsim 1$ for the kinetic \Alfven wave, where Hall MHD predicts  a 
quantitatively different real frequency from kinetic theory, all of
the differences between Hall MHD and kinetic theory occur when the
damping becomes strong, with $\gamma_v/\omega_v> 0.1$. That the weakly
damped wave modes should be well represented by Hall MHD is precisely
what is expected---the fluid limit of the kinetic plasma
physics is valid only when kinetic effects, such as collisionless
damping by wave-particle interactions, are negligible
\citep{Ballai:2002}.

\section{Discussion}
\label{sec:disc}
The detailed comparison of linear eigenfrequencies presented in
\secref{sec:results} provides a solid quantitative foundation upon
which to construct a thorough discussion of the central question of
this paper: \emph{How does choosing the standard Hall MHD system
affect the description of the dynamics and evolution of turbulence in
a weakly collisional plasma?}

One functional definition of turbulence is that it is nothing more
than the sum of all nonlinear wave-wave couplings that serve to
transfer energy from one spatial scale to another; in a three-dimensional
physical system such as the solar wind, the generally accepted picture
is that energy in large scale fluctuations is transferred via these
nonlinear interactions to ever smaller scales, cascading until
eventually reaching a  scale at which some dissipation mechanism
damps the fluctuations, ultimately thermalizing their energy (see
\citet{Schekochihin:2007} and \citet{Howes:2008c} for a theoretical
picture of how this process may occur in a weakly collisional
plasma). Three important questions essentially frame the frontier of
research on kinetic turbulence, and are particularly relevant to an
evaluation of the Hall MHD description for turbulence in a kinetic
plasma:
\begin{enumerate}
\item What are the characteristic wave modes of the turbulence at scales
$k d_i \gtrsim 1$?
\item Is the collisionless damping of the turbulence at  scales $k d_i \gtrsim 1$
well described by linear damping rates?
\item Are the nonlinear wave-wave couplings that drive the turbulent cascade
altered by the presence of unphysical wave modes?
\end{enumerate}
\subsection{The Wave Modes Comprising Kinetic Turbulence}
By exploiting the highly super-Alfv\'enic speed of the solar wind
flow, the temporal fluctuations measured by \emph{in situ} spacecraft
can be mapped, according to the Taylor hypothesis \citep{Taylor:1938},
to spatial fluctuations. Thus, the break in the solar wind magnetic
energy spectrum, typically observed at around $f\sim 0.4$~Hz, is
thought to correspond to some characteristic length scale in the plasma,
perhaps the ion inertial length $d_i$ or the ion Larmor radius
$\rho_i$ \citep{Leamon:1998a,Howes:2008b,Hamilton:2008}. At
frequencies above this spectral break, corresponding to yet smaller
spatial scales, the magnetic energy spectrum appears steeper
\citep{Smith:2006}---this regime is denoted the ``dissipation range.''

The kinetic theory results presented in \secref{sec:kpkz} clearly
demonstrate that, for a weakly collisional plasma with a finite ion
temperature---such as the solar wind---all slow branch modes, the ion
cyclotron wave of the \Alfven branch, and the fast branch ion
Bernstein wave at $k_\parallel d_i \gtrsim 0.1$ are strongly
damped. Only three weakly damped electromagnetic wave modes exist at
scales $k d_i \gtrsim 1$: the kinetic \Alfven wave, the parallel
whistler wave, and the oblique whistler wave. It seems unlikely, even
in a nonlinearly turbulent plasma, that wave modes with strong linear
kinetic damping rates could be responsible for the dissipation range
fluctuations.  The linear wave properties are quite likely to play a
large role in the turbulent dynamics, in particular at the small
length scales $k d_i \gtrsim 1$ where the amplitude of the magnetic
fluctuations is small compared to the magnitude of the local
interplanetary magnetic field \citep{Howes:2008b}. This line of
reasoning suggests that the fluctuations in the dissipation range must
consist of one, or perhaps a mixture, of the three undamped
electromagnetic wave modes.  The finding of \secref{sec:bt} suggests
that, for typical solar wind parameters with $\beta_i \lesssim 1$,
Hall MHD generally does an adequate job of describing these undamped
wave modes.

\subsection{Kinetic Damping Rates}
How effective are linear collisionless damping mechanisms in a
turbulent plasma where nonlinear interactions are continually
transferring energy from one wave mode to another?  This question
essentially weighs the importance of linear kinetic wave-particle
interactions versus nonlinear fluid wave-wave interactions in
turbulent, weakly collisional plasma. According to models for
turbulent fluids, the energy in a particular wave mode will be
transferred to other wave modes through nonlinear interactions on some
characteristic nonlinear time scale. The theory for strong
incompressible MHD turbulence takes this nonlinear timescale to be the
eddy-turnover time in the perpendicular direction
\citep{Goldreich:1995}; the central conjecture of this theory is that,
as the turbulent energy cascades to smaller scales, this nonlinear
timescale remains in an approximate critical balance with the linear
\Alfven wave period, and consequently predicts an anisotropic cascade to high
perpendicular wavenumbers that is supported by numerical evidence
\citep{Cho:2000,Maron:2001} and is consistent with a careful analysis
of solar wind turbulence observations \citep{Horbury:2008}.

The key question in weakly collisional plasmas is whether or not this
nonlinear transfer suppresses the linear collisionless damping via
wave-particle interactions. Conversely, it is possible that nonlinear
wave-wave interactions can transfer energy into a strongly damped
mode, ultimately leading to energy loss from the turbulent
fluctuations through linear wave-particle interactions of that
strongly damped mode. But it is conceivable that nonlinear couplings
involving strongly damped modes are inhibited by an impedance
mismatch, thus preventing energy loss in such a manner. Ultimately, of
course, the turbulence must be dissipated via kinetic damping
mechanisms \citep{Howes:2008c}; in this case, an evaluation of both
the rate of energy transfer by the turbulent cascade and the rate of
kinetic damping is necessary to determine the dissipation as a
function of wavenumber, as in the cascade model of
\citet{Howes:2008b}.

These important questions will most likely only be answered through
detailed nonlinear kinetic numerical simulations. On this front,
turbulence simulations using undamped fluid models, such as Hall MHD,
will provide valuable comparison points for analysis and
interpretation of the kinetic simulation results.

Until such work can provide further guidance, it seems a reasonable
strategy to assume the applicability of the linear kinetic damping
rates to turbulent plasmas. In fact, the first such nonlinear kinetic
simulation of turbulence at the scale of the ion Larmor radius has
been well fit by a model assuming linear damping rates
\citep{Howes:2008a}. The plasma parameters chosen for this study,
however, predicted rather weak linear collisionless damping over the
range of scales simulated, so we must await further simulations in
more strongly damped parameter regimes to test more thoroughly the
effectiveness of the kinetic damping in a turbulent plasma. The study
did show, however, that the damping via kinetic mechanisms in the
turbulent plasma was not stronger than the linear prediction, a result
that is not obvious \emph{a priori}.

Despite the many unanswered questions about damping in kinetic
turbulence, there is no question that the dissipation of turbulence in
a weakly collisional plasma \emph{cannot} be studied by Hall MHD, or
any standard fluid model. The viscous and resistive dissipation terms
used in such models are merely \emph{ad hoc} fluid closures that do
not accurately represent the underlying kinetic mechanisms.  Only a
much more complicated model, such as a Landau-fluid model
\citep{Snyder:1997,Passot:2007}, is capable of adequately representing
the kinetic dissipation. The steeper spectral index of the magnetic
energy spectrum in the dissipation range of solar wind turbulence has
been variously attributed to either kinetic dissipation
\citep{Coleman:1968,Leamon:1999,Gary:1999,Howes:2008b} or to 
wave dispersion \citep{Stawicki:2001,Krishan:2004,Galtier:2006}.  If
dissipation does play a significant role, further investigation
requires a kinetic model. It has even been argued that, without a
kinetic model to determine that kinetic dissipation is negligible for
a specific case, the use of a fluid model such as Hall MHD is
unjustified \citep{Howes:2008d}.

\subsection{Nonlinear Wave-Wave Interactions}

Hall MHD has proven to be a valuable framework for the study of a
number of plasma phenomena, in particular magnetic reconnection. For
example, the finding that whistler waves mediated a faster
reconnection rate \citep{Mandt:1994} paved the way in identifying the
importance of the Hall term in magnetic reconnection. Such
applications depend on the accurate description of the linear
properties of a particular wave mode. But turbulence is consequence of
nonlinear wave couplings; to provide a useful framework for
kinetic turbulence, a model must accurately describe the behavior not of
just one mode, but of all modes.\footnote{The nonlinear interactions in a
turbulent plasma will depend not only on the complex eigenfrequencies
of the normal modes but on their eigenfunctions as well. Although this
paper restricts its focus to the frequencies, a detailed comparison of
all mode properties at $k d_i =0.1$ found that the kinetic mode
properties are not always well represented by Hall MHD \citep{Krauss-Varban:1994}.}

Because the turbulent dynamics and evolution will depend on the
nonlinear couplings between all possible wave modes, the existence of
certain undamped wave modes in Hall MHD---modes that are strongly
damped in a weakly collisional plasma according to kinetic theory---is
troubling. These spurious wave modes effectively provide additional
degrees of freedom to the turbulence that would otherwise be strongly
impeded in a kinetic plasma. Consider, for example, waves in the
parameter regime $k_\parallel d_i \gtrsim 1$ in a plasma with
$\beta_i=1$ and $T_i/T_e=1$ as presented in \figref{fig:kpar_warm} of
\secref{sec:fti}. In Vlasov-Maxwell kinetic theory, both the \Alfven
and slow branch modes are heavily damped in this regime, leaving only
fast branch waves available. But the Hall MHD model supports undamped
\Alfven and slow waves.  It seems unlikely that the nonlinear dynamics
would be the same in the presence of three undamped modes as when only
the fast mode is available, but just how the presence of the
unphysical modes will alter the nonlinear wave-wave couplings is
unclear. Here, once again, a promising avenue for progress is a
detailed study contrasting nonlinear couplings in the presence and
absence of these spurious modes.  A suite of nonlinear kinetic
simulations compared to Hall MHD simulations in the same regime will
provide valuable insight, identifying the regimes of validity for
which Hall MHD provides an adequate description.

Two arguments exist that may serve to diminish the impact of the
spurious undamped waves in Hall MHD. First, as shown in
\figref{fig:kperp_warm}, in the limit that $k_\perp \gg k_\parallel$,
the fast wave frequency is much higher than the \Alfven wave
frequency; since the strength of nonlinear interactions typically
diminishes rapidly as the wave mode frequencies become widely
separated, the presence of a spurious wave mode may negligibly impact
the turbulent couplings in such a case. Of course, in the opposite
limit $k_\perp \ll k_\parallel$, seen in
\figref{fig:kpar_warm}, all three wave modes have similar frequencies for
$k_\parallel d_i \lesssim 1$. A second argument, derived in the
gyrokinetic limit\footnote{The fast wave branch is ordered out of the
system in the gyrokinetic approximation.}  $k_\perp \gg k_\parallel$
and $\omega \ll \Omega_i$, shows that the turbulent cascade of \Alfven
waves does not exchange energy with the slow wave cascade except in
the regime where the perpendicular scale is near the ion Larmor radius
$k_\perp
\rho_i \sim 1$ \citep{Schekochihin:2007}.  Thus, the
existence of a spuriously undamped slow wave may not influence the
\Alfven wave cascade. To explore the exchange of energy between the
separate cascades at the scale of the ion Larmor radius $k_\perp
\rho_i \sim 1$, however, will certainly require nonlinear kinetic
simulations.

\conclusions
\label{sec:conc}
As a model that extends beyond the limits of MHD, Hall MHD has seen
increasing use in recent years as a framework for describing
turbulence in weakly collisional plasmas, such as the near-earth solar
wind. Its applicability to turbulence in kinetic systems has been
called into question \citep{Krauss-Varban:1994,Howes:2008d}, so a
thorough evaluation of the limitations of Hall MHD in this context is
desirable.  This paper takes the first step in this process by
quantitatively comparing the real linear eigenfrequencies of standard
Hall MHD with the complex linear eigenfrequencies of Vlasov-Maxwell
kinetic theory.

Previous work has shown that Hall MHD is a rigorous limit of kinetic
theory only in the cold ion limit satisfying $T_i \ll T_e$, and
$k_\parallel v_{ti} \ll \omega \ll k_\parallel v_{te}$
\citep{Ito:2004,Hirose:2004}; the fluid description is only valid when
wave-particle interactions and finite-Larmor-radius effects are
negligible \citep{Ballai:2002}.  The quantitative comparisons with
kinetic theory in \secref{sec:results} bear out the general finding
that Hall MHD is a valid limit of Hall MHD in the cold ion limit $T_i
\ll T_e$; for finite ion temperature $T_i \sim T_e$, however, the lack
of collisionless damping in Hall MHD leads to undamped wave modes that
do not exist in a weakly collisional plasma.

Three key issues are identified regarding the use of Hall MHD to
describe turbulence in kinetic plasmas: (1) what are the wave modes
comprising the turbulence at scales $k d_i \gtrsim 1$?; (2) are the
collisionless damping rates from linear kinetic theory applicable in a
nonlinearly turbulent plasma?; and (3) are the nonlinear wave-wave
mode couplings inherent in turbulence altered by the presence of
spurious, undamped wave modes in Hall MHD? In a weakly collisional
plasma, the only three undamped electromagnetic wave modes that exist
at scales $k d_i \gtrsim 1$ are the kinetic \Alfven wave, the parallel
whistler wave, and the oblique whistler wave; each of these waves is
generally well-described by Hall MHD for $\beta_i \lesssim 1$. To
determine the effective collisionless damping in a turbulent plasma, a
research program using nonlinear numerical simulations to contrast the
predictions of a kinetic approach with those of a fluid approach is
the most promising path forward.  For studies focusing on the
dissipation of turbulence and the thermalization of the turbulent
energy, a kinetic description is certainly required.  Both fluid Hall
MHD and kinetic nonlinear numerical simulations will be instrumental
in shedding light on the question of whether the presence of the
spuriously undamped waves in Hall MHD alters the nonlinear couplings
of the available wave modes, thus changing the dynamics and evolution
of the turbulence due to unphysical effects.  A study focused on the
turbulent dynamics at the characteristic length scales in the plasma,
such as the ion Larmor radius or the ion inertial length, is the
natural starting point for such a numerical investigation.  In
conclusion, although nonlinear kinetic simulations will be
indispensable for the study of turbulence at kinetic scales in weakly
collisional plasmas, Hall MHD will certainly continue to provide
useful insights and valuable points for comparison in the study of
kinetic turbulence.

\begin{acknowledgements}
The author thanks Ben Chandran for insightful discussions.
\end{acknowledgements}


\end{document}